\documentclass[journal,comsoc,numbers,sort&compress]{IEEEtran}
\usepackage{cite}
\usepackage{graphicx}
\usepackage{epstopdf}
\usepackage{enumitem}
\usepackage{subfigure}
\usepackage{amssymb, amsmath}
\usepackage{amsthm}

\usepackage{makecell,multirow,diagbox}
\usepackage{stfloats}
\usepackage{array}
\usepackage{stfloats}
\usepackage{colortbl} 
\usepackage{etoolbox}
\usepackage{xtab,booktabs}
\usepackage{rotating}
\usepackage{enumerate}

\begin{document}	
\title{Game Theoretic Approaches in Vehicular Networks: A Survey}
	\author{
		Zemin~Sun,~\IEEEmembership{} 
		Yanheng~Liu,~\IEEEmembership{}
		Jian~Wang,~\IEEEmembership{Member, IEEE}
		Carie~Anil,~\IEEEmembership{}
		Dongpu~Cao,~\IEEEmembership{Senior Member, IEEE}
		\thanks{ Z.M. Sun, Y.H. Liu, and J. Wang are with the College of Computer Science and Technology, and Key Laboratory of Symbolic Computation and Knowledge Engineering of Ministry of Education, Jilin University, Changchun 130012, China (e-mail: laurasun166@gmail.com, yhliu@jlu.edu.cn, wangjian591@jlu.edu.cn).}
		\thanks{ C. Anil is with the School of Computer Science and	Engineering, VIT-AP, Amaravati-522237, Andhra Pradesh, India, and the College of Engineering, Nanjing Agricultural University, Nanjing 210031, China.(email:carieanil@gmail.com) }
		\thanks{ D.P. Cao is with Waterloo Cognitive Autonomous Driving Lab, University of Waterloo, N2L 3G1, Canada (e-mail: dongpu.cao@uwaterloo.ca). Zemin Sun is also with this affiliation.}

		\thanks{Manuscript received,  revised .}
}
	\markboth{Journal of \LaTeX\ Class Files,~Vol.~13, No.~9, September~2014}
	{Shell \MakeLowercase{\textit{et al.}}: Bare Demo of IEEEtran.cls for Computer Society Journals}
	
	\maketitle
	\begin{abstract}
In the era of the Internet of Things (IoT), vehicles and other intelligent components in Intelligent Transportation System (ITS) are connected, forming the Vehicular Networks (VNs) that provide efficient and secure traffic, ubiquitous access to information, and various applications. However, as the number of connected nodes keeps increasing, it is challenging to satisfy various and  large amounts of service requests with different Quality of Service (QoS ) and security requirements in the highly dynamic VNs. Intelligent nodes in VNs can compete or cooperate for limited network resources so that either an individual or a group objectives can be achieved. Game theory, a  theoretical framework designed for strategic interactions among rational decision-makers who faced with scarce resources, can be used to model and analyze individual or group behaviors of communication entities in VNs.  This paper primarily surveys the recent advantages of GT used in solving various challenges in VNs. As VNs and GT have been extensively investigate34d, this survey starts with a brief introduction of the basic concept and classification of GT used in VNs. Then, a comprehensive review of applications of GT in VNs is presented, which primarily covers the aspects of QoS and security. Moreover, with the development of fifth-generation (5G) wireless communication, recent contributions of GT to diverse emerging technologies of 5G integrated into VNs are surveyed in this paper. Finally, several key research challenges and possible solutions for applying GT in VNs are outlined.

\end{abstract}
	
	\maketitle

	\IEEEpeerreviewmaketitle

\section{Motivations and Objectives}
\label{sec:introduction}
	
 Internet of things (IoT) is seen as the most promising technology to realize the vision of connecting things at any time, from any place to any network, and continues to be a hot pot of the research. Integrating IoT technologies to vehicles and infrastructures, Intelligent Transportation System (ITS) aims to improve traffic safety, relieve traffic congestion and increase energy efficiency by providing real-time information for road users and transportation system operators \cite{ni2015traffic}. Vehicular networks (VNs) has been regarded as an important component of the development of the ITS for intelligent decision makings by communicating with other vehicles (V2V) or infrastructures (V2I) \cite{gasmi2019vehicular}. 
 
 However, the salient characteristics of VNs such as high-mobility of vehicles, unstable network topology, and constraint resources (e.g., power, frequency spectrum, and bandwidth) pose severe challenges to decision makings of different intelligent devices (e.g., vehicles and Road-Side Units (RSUs)) in dynamic environments. For example, RSUs make decisions on resource allocation to improve resource utilization while minimizing computational overheads and energy consumption. Besides, resource consumers such as vehicles make decisions to either selfishly compete for limited resources or cooperate with others to maximize the overall network performance. Moreover, to make the communication secure against the selfish behaviors or attacks of malicious nodes, intelligent devices in VNs should make efficient defense decisions. Game theory (GT) \cite{friedman1986game} provides mathematical models for the optimization of complex issues where multi-players with contradictory objectives compete for limited resources or cooperate for maximizing common interests. Therefore, this paper provides a systematic survey on applications of GT in modeling the strategic behaviors of intelligent nodes in VNs.

 VNs have been extensively studied in different contexts, and the recently published surveys cover many aspects of the VNs. Several surveys focus on the Quality-of-Service (QoS) of VNs involving the protocols of Medium Access Control (MAC) \cite{jayaraj2016survey} and network layers \cite{peng2018vehicular}, routing algorithms \cite{awang2017routing}, content delivery solutions \cite{silva2016vehicular}, and energy harvesting techniques \cite{atallah2016energy}. Furthermore, considering the vulnerability of VNs to various attacks, several research investigate security \cite{bariah2015recent}, privacy \cite{manivannan2020secure,boualouache2017survey}, and trust \cite{kerrache2016trust} aspects of VNs. The researchers in \cite{lu2018survey} gave a comprehensive overview covering the fundamentals of VNs and the three security-related topics: security, privacy, and trust. With the development of the proposed fifth-generation (5G) communication architecture \cite{shah20185g}, some surveys targeted at the specific technologies integrating with VNs including vehicular edge computing (VEC)\cite{liu2019vehicular,dziyauddin2019computation}, big-data-driven VNs \cite{cheng2018big}, Vehicular cloud networks \cite{mekki2017vehicular}, unmanned aerial vehicles (UAV)-assisted VNs \cite{shi2018drone}, and  heterogeneous vehicular networks (HVNETs) \cite{zheng2016soft,zheng2015heterogeneous}. However, these papers mainly focus on characteristics, requirements, attacks, and corresponding solutions for VNs but with little coverage on challenges of decision making mentioned above in VNs.

Although GT has been used with significant success to design the interactions of competitive and cooperative behaviors among players, existing survey papers on GT focus mostly on either the overview \cite{oulaourf2017review,moura2017survey} and security aspects \cite{sedjelmaci2019cyber,pawlick2019game,do2017game} of wireless networks or single technology-based wireless networks \cite{mkiramweni2019survey,moura2018game}. For example, a survey of GT based solutions for radio resource allocation optimization in fourth-generation (4G) wireless networks is presented in \cite{oulaourf2017review}. GT applying in solving challenges in emerging wireless communications such as Mobile Edge Computing (MEC), Device to Device (D2D), Cognitive Radio VNs (CR-VN), and sensor networks are summarized in \cite{moura2017survey}. Moreover, surveys in \cite{mkiramweni2019survey} and \cite{moura2018game} uniquely focus on the game-theoretic approaches for dealing with the issues in UAV-assisted wireless networks and multi-access edge computing wireless networks, respectively. Besides, several survey papers address the application of GT in decision making of cyber security and privacy protection in wireless networks\cite{sedjelmaci2019cyber,pawlick2019game,do2017game}. Although GT approaches in VNs is introduced in \cite{bahamou2016game}, the paper mainly gives a brief introduction of the GT employed for modeling the interaction between the attacker and defender in VNs. Little survey work has been carried out comprehensively for applications of GT in VNs. The classification of  related surveys and the research gap are shown in Table. \ref{tab_RelatedSurveys}.

The focus of this paper is to provide a comprehensive survey of the current research on GT approaches in solving the challenges facing VNs. The main contributions can be summarized as: 
\begin{itemize}
	\item Providing a comprehensive overview of the existing applications of  GT in improving the performance of VNs from the perspectives of QoS guarantee and security protection.
	
	\item Discussing the GT applied to solve the challenges in next-generation VNs integrating with  5G-based technologies such as edge computing, heterogeneous networks, UAV-assisted communications, and SDN technologies. 
	
	\item Outlining the remaining challenges and future researches that help to advance the development of GT in VNs.

\end{itemize}

\begin{figure*}[!hbt]
	\centering
	\includegraphics[width =6in]{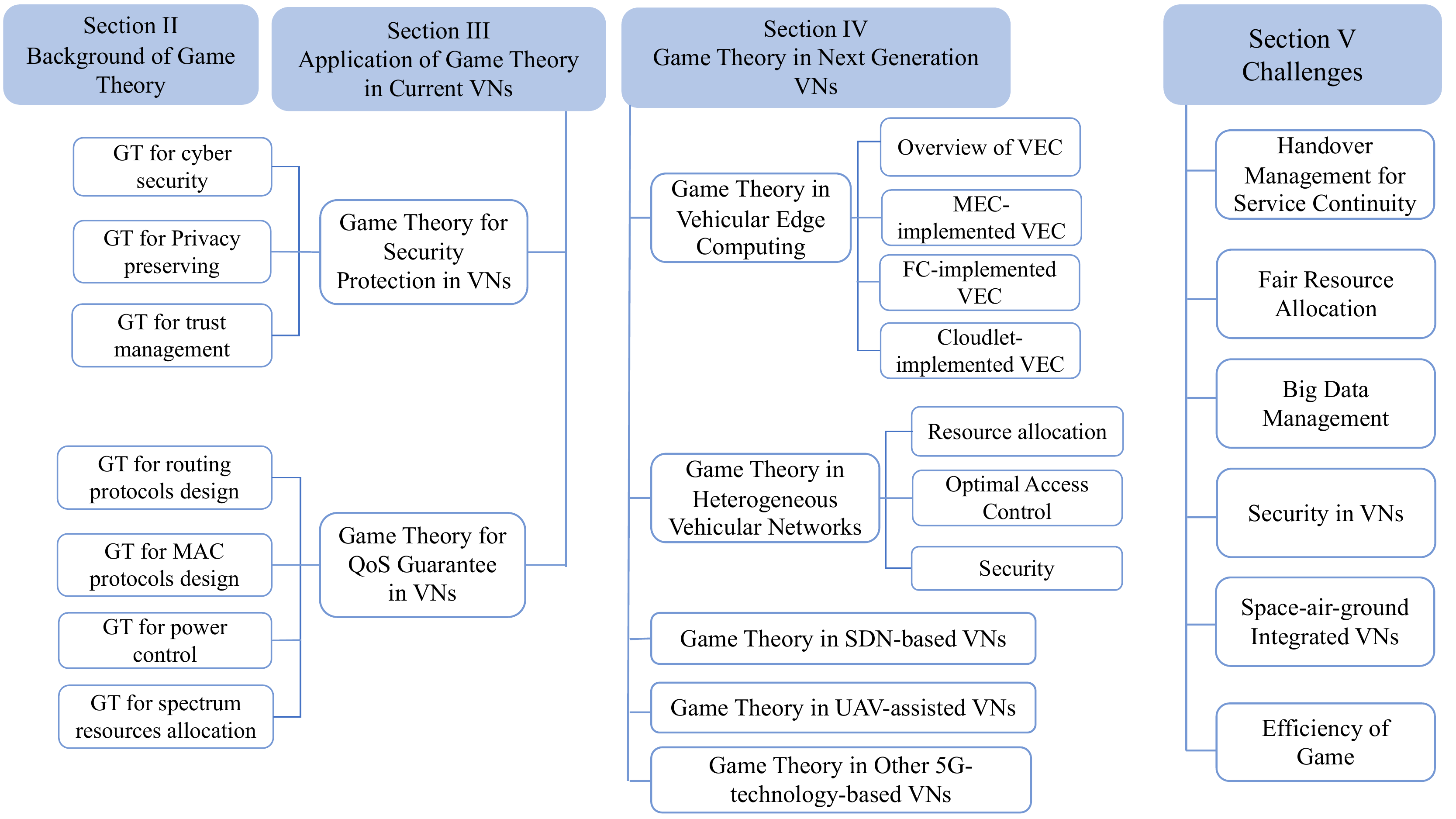}
	\caption{Organization of the survey on applications of GT in VNs}
	\label{fig_outline}
\end{figure*}

Fig. \ref{fig_outline} illustrates the organization of this survey. The rest of the paper is organized as follows. Section \ref{sec_background} briefly introduces the background of GT. Section \ref{sec_GTinVNs} reviewed and organized the GT solutions for the challenges of current VNs. Section \ref{sec_GTin5G} further investigates the applications of GT in VNs with 5G-based technologies. Section \ref{sec_challenge} highlights the key challenges and future works. The paper is concluded in Section \ref{sec_conclusion}. The abbreviations in this paper are listed in Table \ref{tab_abb}.

	\begin{table}
	\scriptsize
	\caption{List of Acronyms and Corresponding Definitions}
	\label{tab_abb}
	\renewcommand*{\arraystretch}{1}
	\begin{center}
		\begin{tabular}{p{.06\textwidth} p{.28\textwidth}}
			\textbf{Symbols} &\textbf{Description}\\
			\hline
			IoT&Internet of Things\\
			ITS &Intelligent Transportation Systems\\
			VN&Vehicular Networks\\
			V2V &Vehicle to Vehicle\\
			V2I &Vehicle  to Infrastructure\\
			RSU&Road-Side Unit\\
			GT&Game Theory\\
			5G&	Fifth Generation \\
			QoS&Quality of Service\\
			MAC&Medium Access Control\\ 
 	        4G& Fourth Generation  \\
 	        MEC&Mobile Edge Computing\\
 	        D2D&Device to Device\\
			NE&Nash Equilibrium\\
			BNE&Bayesian Nash Equilibrium\\
			DoS&Denial of Service\\
			IDS&Intrusion Detection System\\
			IPS& Intrusion Prediction System\\
			IRS& Intrusion Reaction System \\
			CH&Cluster Head\\
			CH-IDS&Coalition Head with Intrusion Detection Head\\
			LIDS&Local Intrusion System Detection\\
			GDS&Global Decision System\\
			GIDS&Global Intrusion Detection System\\
			IDA&Intrusion Detection Agent\\
			 PKI& Public Key Infrastructure\\
			PKI&Public Key Infrastructure\\
	        LA&Learning Automata\\
	        NBS&Nash Bargaining Solution\\	        
	        CR&  Cognitive Radio\\
	        MANETs& mobile ad hoc networks\\
	        SNR& Signal-to-Noise Ratio\\
	        V2R&  Vehicle-to Roadside\\
	        TDMA&Time Division Multiple Access\\
	         CSMA&Carrier Sense Multiple Access\\
	        CSMA/CA&Carrier Sense Multiple Access/Collision Avoidance\\
	        BSM&Basic Safety Message\\
	        CBR&Channel Busy Ratio\\
	        DCC& Decentralized Congestion Control\\
	        SU&Secondary User\\
	        PU&Primary User\\
			SINR&Signal-to-Interference-plus-Noise-Ratio\\
		    MEC&Mobile Edge Computing\\
			HetVNET&Heterogeneous Vehicular Network\\
			SDN&Sofware Defined Network\\
			UAV&Umanned Assisted Vehicles\\
			VEC&Vehicular Edge Computing\\
			VC&Vehicular Cloud\\
			FC&Fog Computing\\
			VFC&Vehicular Fog Computing\\
			CC&Cloud Computing\\
		    SP&Service Provider\\
		    VM&Virtual Machine\\
		    QoE&Quality of Experience\\
		    ECD&Edge Computing Device\\
		    VFN&Vehicular Fog Node\\
		    F-SBS&Fog-Small-BS\\
			HetVNETs& Heterogeneous Vehicular Networks\\
			LTE&Long Term Evolution\\
			DSRC&Dedicated short-range communication \\
			ROS&Requirement of Service\\
			CP&Content Provier\\
			MNO&Multiple Network Operators\\
			OBUs &On Board Units \\
			V2U&Vehicle-to-UAV\\
			HP & High Priority \\
			LP&Low Priority\\
			mmWave& Millimeter-wave (mmWave)\\
			ISPs& Internet Service Providers \\
			LSA& Licensed Shared Access\\
			RL&Reinforce Learning\\
		\end{tabular}
	\end{center}
\end{table}

\begin{table*}
	\caption{Classification of related surveys}
	\footnotesize
	\label{tab_RelatedSurveys}
	\renewcommand*{\arraystretch}{1.2}
	\begin{center}
	\begin{tabular}{|p{.5\textwidth}|p{.4\textwidth}|p{.028\textwidth}|}
			\hline 
			      \textbf{Title} &\textbf{Published}&\textbf{Year}\\ 
			\hline
				  \multicolumn{3}{|c|}{\textbf{VN Related Surveys}}\\
		    \hline
			   Secure authentication and privacy-preserving techniques in Vehicular Ad-hoc NETworks (VNs) \cite{manivannan2020secure} &Vehicular Communications&2020\\
		    \hline
		       Vehicular edge computing and networking: A  survey \cite{liu2019vehicular}&arXiv&2019\\
		       Computation offloading and content caching delivery in vehicular edge computing: A survey \cite{dziyauddin2019computation}&arXiv preprint arXiv:1912.07803&\\
		       \hline
		        5G for vehicular communications \cite{shah20185g}&IEEE Communications Magazine&2018\\
		    	Vehicular communications: A network layer perspective \cite{peng2018vehicular}&IEEE Transactions on Vehicular Technology&\\
		    	A survey on recent advances in vehicular network security, trust, and privacy \cite{lu2018survey}&IEEE Transactions on Intelligent Transportation Systems&\\
		    	Big data driven vehicular networks \cite{cheng2018big}&IEEE Network&\\
		    	Drone assisted vehicular networks: Architecture, challenges and opportunities \cite{shi2018drone}&IEEE Network&\\
		   \hline
		          A survey on pseudonym changing strategies for vehicular ad-hoc networks \cite{boualouache2017survey}&IEEE Communications Surveys \& Tutorials&2017\\
		         Routing in vehicular ad-hoc networks: A survey on single-and cross-layer design techniques, and perspective \cite{awang2017routing}&IEEE Access&\\
		         Vehicular cloud networks: Challenges, architectures, and future directions \cite{mekki2017vehicular}&Vehicular Communications&\\
		   \hline  
		          Soft-defined heterogeneous vehicular network: Architecture and challenges \cite{zheng2016soft}&IEEE Network&2016\\
		          Vehicular networks: A new challenge for content-delivery-based applications \cite{silva2016vehicular}&ACM Computing Surveys (CSUR)&\\
		          Energy harvesting in vehicular networks: A contemporary survey \cite{atallah2016energy}&IEEE Wireless Communications&\\
		          Trust management for vehicular networks: An adversary-oriented overview \cite{kerrache2016trust}&IEEE Access&\\
		          A survey on hybrid MAC protocols for vehicular ad-hoc networks \cite{jayaraj2016survey}&Vehicular communications&\\
		      \hline     
		          Heterogeneous vehicular networking: A survey on architecture, challenges, and solutions \cite{zheng2015heterogeneous}&IEEE communications surveys \& tutorials&2015\\
		   \hline
		        \multicolumn{3}{|c|}{\textbf{GT Related Surveys}}\\
		   \hline
		        A survey on the combined use of optimization methods and game theory \cite{sohrabi2020survey}&Archives of Computational Methods in Engineering&2020\\
		 \hline 
		        Cyber security game for intelligent transportation systems \cite{sedjelmaci2019cyber}&IEEE Network&2019\\
		       A game-theoretic taxonomy and survey of defensive deception for cybersecurity and privacy \cite{pawlick2019game}&ACM Computing Surveys (CSUR)&\\
		        A Survey of game theory in unmanned aerial vehicles communications \cite{mkiramweni2019survey} &IEEE Communications Surveys \& Tutorials&\\
	       \hline    
	            Game theory for multi-access edge computing: Survey, use cases, and future trends \cite{moura2018game}&IEEE Communications Surveys \& Tutorials&2018\\
            \hline    
		         Survey of game theory and future trends with application to emerging wireless data communication networks \cite{moura2017survey}&Researchgate. net/publication&2017\\
		        Game theory for cyber security and privacy \cite{do2017game}&ACM Computing Surveys (CSUR)&\\
		        Review on radio resource allocation optimization in LTE/LTE-advanced using Game Theory \cite{oulaourf2017review}&International Journal of Communication Networks and Information Security (IJCNIS)&\\
	            \hline  
	           \multicolumn{3}{|c|}{\textbf{GT for Security of VNs}}\\
	            \hline  
		       When GT meets VN's security and privacy \cite{bahamou2016game}&Proceedings of the 14th International conference on advances in mobile computing and multi media&2016\\
				 \hline  
				\multicolumn{3}{|c|}{\textbf{GT in VNs}}\\
				\hline  
		         Game theory approaches in vehicular networks: A survey& \multicolumn{2}{c|}{This paper}\\
		    \hline  	        
    \end{tabular}
	\end{center}
\end{table*}

\section{Application of game theory in current VNs}
\label{sec_GTinVNs}
 As GT has been extensively investigated, this section starts with a brief introduction of the basic concept and classification of GT used in VNs.  Furthermore, related researches on applications of  GT in VNs are introduced from the perspectives of security protection and QoS guarantee, respectively. 
\subsection{Background of Game theory}
\label{sec_background}

GT is a branch of applied mathematics for analyzing the strategic interactions among multiple decision-makers (players) \cite{friedman1986game}. These decision-makers cooperatively or competitively take rational actions that have conflicting consequences. A classical game can be defined as a triplet $ \mathbf{G}=(\mathcal{N},( \mathcal{S}_i)_{i\in\mathcal{N}}, (\mathcal{U}_i)_{i\in\mathcal{N}})$, where:
\begin{itemize}
	\item $\mathcal{N}={1,...N}$ is a  finite set of $N$ players 
	\item $(\mathcal{S}_i)_{i\in\mathcal{N}}$ is the strategy set of player $i$, and $s_i\in\mathcal{S}_i$ is any possible strategy of $i$
	\item $ (\mathcal{U}_i)_{i\in\mathcal{N}}) $ is the utility (or payoff) function of  player $i$.  Each player in the game aims to maximize its payoff function according to other players strategies
\end{itemize}

A strategy profile $s^*$ is a \textit{Nash equilibrium (NE)}, if for each player $i \in \mathcal{N}$
\begin{equation}
\label{eq_NE}
u_{i}\left(s_{i}^{*}, s_{-i}^{*}\right) \geq u_{i}\left(s_{i}, s_{-i}^{*}\right), \forall s_{i} \in \mathcal{S}_i,
\end{equation}
where $s_{-i}^{*}$ is the strategy vector of all players except player $ i$. This is the maximum utility function of player $i$ over the set of all its strategies. In this case, each player obtains the optimal utilities, and none of the users has an incentive to deviate because it can not unilaterally change its strategy to increase his utility.

Classical GT can be classified into two main categories: non-cooperative and cooperative games. Non-cooperative game is most widely-used in resource competition or attacker detection due to the simplicity for problem modeling and solution. Besides, the cooperative game has been applied in the resource allocation or sharing problems in VNs. Because the details of GT have been investigated by a lot of surveys or studies on wireless networks \cite{manshaei2013game,felegyhazi2006game,akkarajitsakul2011game,sun2019non,sedjelmaci2019cyber,pawlick2019game,mkiramweni2019survey,moura2018game,moura2017survey,do2017game,oulaourf2017review,bahamou2016game},  we skip the introduction of GT by summarizing the taxonomies and applications of GT corresponding to this paper in Table \ref{tab_game}.

\begin{table*} 
	\caption{Taxonomy and applications of GT in VNs }
	\label{tab_game}
	\scriptsize
	\renewcommand*{\arraystretch}{1}
	\begin{center}
		\begin{tabular}{|m{.005\textwidth}|m{.004\textwidth}|m{.04\textwidth}|m{.043\textwidth}|m{.04\textwidth}|m{.04\textwidth}|m{.06\textwidth}|m{.04\textwidth}|m{.04\textwidth}|m{.06\textwidth}|m{.06\textwidth}|m{.05\textwidth}|m{.07\textwidth}|m{.05\textwidth}|m{.04\textwidth}|}
			\hline
			\multicolumn{3}{|c|}{\textbf{Games}}&\textbf{Cyber Security}&\textbf{Privacy}&\textbf{Trust}&\textbf{Routing}&\textbf{MAC Layer Access}&\textbf{Power Control}&\textbf{Spectrum Resource Allocation}&\textbf{VEC}&\textbf{HetVNET}&\textbf{SDN-based VNs}&\textbf{UAV-assisted VNs}&\textbf{Others}\\
			\hline
			\multirow{8}{1cm}{\rotatebox[origin=c]{90}{\textbf{Non-cooperative}}}&\multicolumn{2}{c|}{\textbf{General}}&\cite{sedjelmaci2015accurate,subba2018game}&\cite{fan2018network,wang2019optimization} & \cite{mehdi2017game,fan2019trust}&\cite{tian2017self,das2017new,hu2016novel,assia2019game,goudarzi2018non,goudarzi2019fair} &\cite{li2019tcgmac,wang2018application,lang2019vehicle,al2017cooperative}  & \cite{goudarzi2018non,goudarzi2019fair,hua2017game,sun2017non}&&\cite{zhang2019task,zhang2017mobile,zhao2019computation,yu2016optimal,tao2017resource,aloqaily2017fairness,brik2018gss}&\cite{xiao2018spectrum,zhao2019optimal} &&\cite{xiao2018uav,alioua2018efficient} &\\
			\cline{2-15}
			&\multicolumn{2}{c|}{\textbf{Stackelberg}}&&\cite{brahmi2019cyber,sedjelmaci2018generic}&&&&&&&\cite{zhang2017optimal,zhou2018begin,liwang2019game,lin2019vehicle}& \cite{li2016control,aujla2017data,chahal2019network,alioua2019incentive} &&Network slicing \cite{zhou2018bandwidth} \\
			\cline{2-15}
			&\multirow{3}{1cm}{\rotatebox[origin=c]{90}{\textbf{Bayesian}}}&\textbf{General}&\cite{sedjelmaci2014detection,behfarnia2019misbehavior}&\cite{ying2015motivation}&&& \cite{kwon2016bayesian} &&&&&&\cite{sedjelmaci2016intrusion} &\\
			\cline{3-15}
			&&\multirow{2}{1cm}{\textbf{Signaling}}&&&\multirow{2}{1cm}{\cite{haddadou2014job}}&&&&&\multirow{2}{1cm}{\cite{hui2019edge}}&\multirow{2}{1cm}{\cite{mabrouk2016meeting} }&&&\\
			&&&&&&&&&&&&&&\\
			\cline{2-15}
			&\multicolumn{2}{c|}{\textbf{Auction}}&&&&&&&\cite{kumar2016spectrum}&&&&&\\
			\cline{2-15}
			&\multicolumn{2}{c|}{\textbf{One-shot}}&&&&\cite{suleiman2017adaptive}&&&&&&&&\\
			\cline{2-15}
			&\multicolumn{2}{c|}{\textbf{Mean Field}}&&&& \cite{zhang2017resource}&&&&&&&&\\
			\cline{2-15}
			&\multicolumn{2}{c|}{\textbf{Congestion}}&&&&&&&&&\cite{chen2017congestion} &&&LSA  \cite{belghiti20185g}\\
			\hline
			\multirow{4}{1cm}{\rotatebox[origin=c]{90}{\textbf{\textbf{Cooperative}}}}&\multicolumn{2}{c|}{\textbf{General}}&& &&&& \cite{shah2018shapely}&&&&&&\\
			\cline{2-15}
			&\multicolumn{2}{c|}{\textbf{Coalitional}}&\cite{mabrouk2018signaling}&\cite{boudagdigue2018distributed} &	\cite{halabi2019trust,kumar2015intelligent} & \cite{wu2018computational,sulistyo2019coalitional,assia2019game} &&&&\cite{yu2015cooperative}&\cite{hui2017optimal,hui2019game} &&&\\
			\cline{2-15}
			&\multicolumn{2}{c|}{\textbf{Bargain}}&& &&\cite{kim2016timed} && \cite{chen2016information}& \cite{eze2019design} & \cite{huang2017distribute}&&&&\\
			\cline{2-15}
			&\multicolumn{2}{c|}{\textbf{Potential}}&& &&&& &&	\cite{liu2018computaion,klaimi2018theoretical}  &&&&\\
			\hline
			\multicolumn{3}{|c|}{\textbf{\textbf{Evolutionary}}}&&&\cite{tian2019evaluating}&\cite{khan2018evolutionary}&&&&  \cite{mekki2017proactive}&& \cite{jia2019bus} &\cite{wang2018mode} &\\
			\hline
			\multicolumn{3}{|c|}{\textbf{\textbf{Matching}}}&&&&&&&\cite{shattal2018channel,tian2019channel} & \cite{gu2019task,xu2018low,zhou2019computation}&&&&D2D \cite{gu2016exploiting}, mmWave\cite{perfecto2017millimeter} \\
			\hline
		\end{tabular}
	\end{center}
\end{table*}

	\subsection{Game Theory for Security Protection in VNs}
	\label{sec_security}
		The features of VNs may result in security vulnerabilities that make the communication suffer from various kinds of internal and external attacks, which have caused three main concerns in the design of secure VNs: security, privacy and trust. This section introduces the use of GT in intrusion or misbehavior detection\cite{mabrouk2018signaling,sedjelmaci2014detection,sedjelmaci2015accurate,subba2018game,brahmi2019cyber,sedjelmaci2018generic}, privacy preservation \cite{fan2018network,wang2019optimization,ying2015motivation,al2018teaming,zhang2017otibaagka,zhang2019pa,mukherjee2019efficient}, and trust management \cite{halabi2019trust,boudagdigue2018distributed,kumar2015intelligent,mehdi2017game,fan2019trust,haddadou2014job,tian2019evaluating} in VNs.
		

	\subsubsection{ GT for cyber security}
	\label{sec_intrusion}
	\begin{table*}
	\scriptsize
	\caption{Game models for cyber security in VNs}
	\label{tab_intrusion}
	\renewcommand*{\arraystretch}{1}
	\begin{center}
		\begin{tabular}{|p{.08\textwidth}|p{.19\textwidth}|p{.3\textwidth}|p{.22\textwidth}|p{.015\textwidth}|p{.015\textwidth}|p{.015\textwidth}|}
			\hline 
			\multirow{2}{1cm}{\textbf{Security games}}&\multirow{2}{2cm}{\textbf{Game type}}&\multirow{2}{3cm}{\textbf{Player and Strategy}}&\multirow{2}{2cm}{\textbf{Solution}}&\multicolumn{3}{c|}{\textbf{Role}}\\
			\cline{5-7}
			&&&&\textbf{IDS}&\textbf{IPS}&\textbf{IRS}\\
			\hline
			\cite{mabrouk2018signaling}& Coalition game \newline Signaling game &\textbf{CH-IDS agent}: $ \left\{Idle, Defend\right\} $  \newline\textbf{ Vehicle member}: $ \left\{Attack, Cooperateright\right\} $&Pure-strategy and mixed-strategy BNEs &$\surd$&&\\
			\hline
			\cite{sedjelmaci2014detection} &Bayesian game&\textbf{ RSU}: $\left\{Detect, Wait \right\}$ \newline \textbf{Attacker}: $ \left\{Attack, Wait \right\}$&Mixed strategy BNE&$\surd$&$\surd$&\\
			\hline
			\cite{sedjelmaci2015accurate} &Non-cooperative game&\textbf{LIDS}  $\left\{Detect,Wait\right\} $ \newline \textbf{vehicles members}: $ \left\{Attack, Be normal\right\} $&Pure strategy NE&$\surd$&&\\
			\hline
			\cite{subba2018game}&Non-cooperative game&\textbf{IDS}: $  \left\{Monitor, Not \ Monitor \right\}$ \newline \textbf{ Malicious vehicle}:$ \left\{Attack, Wait\right\} $&Mixed strategy NE&$\surd$&&\\
			\hline
			\cite{brahmi2019cyber,sedjelmaci2018generic} &Hierarchical Stackelberg game&\textbf{IDA}: launching strategies of the secondary agents\newline \textbf{Secondary agents (IDS,IPS,IRS)}: $ \left\{IDS,IPS,IRS\right\} $&NE&$\surd$&$\surd$&$\surd$ \\
			\hline
			\cite{mejri2016new}&Zero-sum game &\textbf{ Attacker}: $ \left\{Attack,Stop\right\} $\newline \textbf{Honest vehicle}:$  \left\{Continue, Change direction\right\} $&Pure strategy NE&$\surd$&&\\
			\hline	
			\cite{behfarnia2019misbehavior}&Bayesian Game&\textbf{Target node}: $\left\{attack, not attack\right\}$\newline \textbf{Benign player}:$\left\{vote,abstain\right\}$& Pure-strategy and Mixed-strategy BNEs&$\surd$&&\\
			\hline
		\end{tabular}
	\end{center}
\end{table*}

	
   Buinevich et al. \cite{buinevich2019forecasting} listed the top 10 most severe cyber-attacks facing VNs, which are categorized as threats to confidentiality, availability, and integrity. For confidentiality attacks, malicious nodes aim to find the Cluster Head (CH) and its sensitive data by eavesdropping the communication within its radio range \cite{brahmi2019cyber}. Regarding the availability attack (e.g., Denial of Service (DoS)), the adversaries attempt to exhaust the bandwidth resources by disturbing or jamming the communications \cite{sakarindr2007security}. Attacks targeting the communication integrity try to inject false information, change the order of messages, or replay the old messages \cite{ichaba2018security}. Consequently, it is essential to provide a reliable security framework for VNs to deter these cyber attacks.  The frameworks developed for VN cyber security can be classified into Intrusion Detection System (IDS), Intrusion Prediction System (IPS), and Intrusion Reaction System (IRS) \cite{zaidi2015host,bouali2016distributed}. 
      
    IDS or IPS is reliable to detect malicious vehicles and is often used as a second line of defense after the cryptographic systems \cite{erritali2013survey, sedjelmaci2013efficient}. GT can be used as a method for modeling the interaction between malicious nodes and the intrusion decision agent. Moreover, it is proved that GT is useful in increasing the detection or prediction accuracy and decreasing the communication overhead. Table. \ref{tab_intrusion}  classifies the related works on cyber security games for VNs according to the deployed game models and cyber security frameworks.

    Mabrouk et al. \cite{mabrouk2018signaling} introduced a coalition and signaling game-theoretical method for internal malicious nodes detection in VNs. By using the coalitional game \cite{shenoy1979coalition}, vehicles are organized into coalitions, each of which are composed of member vehicles and a Coalition Head that equipped with an Intrusion Detection System (CH-IDS). The interaction between the CH-IDS agent and vehicle member is modeled as a signaling game. The CH-IDS performs the optimal strategy $Idle$, or $Defend$ based on the pure-strategy and mixed-strategy BNEs to defend the malicious vehicles' attack dynamically. This scheme relies only on the current behaviors of attackers in the VNs and cannot predict future misbehaviors.

	 To predict the malicious behavior in VNs, Sedjelmaci et al. \cite{sedjelmaci2014detection} proposed a Bayesian game-theoretic intrusion detection and prevention scheme that can predict the future behaviors of the monitored vehicles. The vehicles within the communication range of the monitored vehicle monitor its behavior when it broadcasts an alert message. The RSU and the suspected vehicle begin to play a Bayesian game if a malicious behavior of the monitored vehicle is detected, t. The RSU could perform \textit{detect} or \textit{wait} and the monitored vehicle could \textit{Attack} or \textit{Wait} (i.e., don't wait). As a result, it is proved that there is a mixed strategy BNE $\left\{\text{RSU} \ \left(Detect ,p^{*}\right), \ \text{Attacker} \ (Attack, q^*)\right\}$ in which the malicious vehicle attacks when the probability $ q > q^*  $ and the RSU triggers its detection action when $ p < p^* $. In this case, the suspected vehicle can be predicted as an attacker and be stored into the blacklist by the RSU. The experiments demonstrated  that this scheme has high rate of detection $(>98\% )$ and low rate of false-positive $ (<2\%)$. The weakness of this method is that it only focuses on detecting the \textit{false alert's generation attack} but does not consider other types of attacks facing the VNs.

    Sedjelmaci et al. \cite{sedjelmaci2015accurate} further developed a lightweight hybrid IDS for VNs considering both the mobility of vehicles and various types of attacks, neither of which were considered in their previous work \cite{sedjelmaci2014detection}. Three intrusion detection agents are considered in the hybrid IDS framework: Local Intrusion Detection System (LIDS) at the cluster member level, Global Intrusion Detection System (GIDS)  at the CH level, and Global Decision System (GDS) at RSU level. At the LIDS level, a game is designed for a cluster member and vehicles located at its radio range; the monitoring process can be activated if the cluster member reaches a NE. This detection scheme outperforms VWCA \cite{daeinabi2011vwca}, IDFV \cite{sedjelmaci2014new}, and T-CLAIDS \cite{kumar2014collaborative} in terms of attack detection accuracy (i.e. detection and false-positive rates),  detection time, and communication overhead.

    There is a trade-off between collecting more information for effective intrusion detection and the high volume IDS traffic in VNs. A multi-layered IDS  for VN based on GT is designed in \cite{subba2018game} to achieve a satisfied tradeoff between QoS and security. The interaction between the IDS and the malicious vehicle is modeled as a two-player non-cooperative game. The strategy spaces of the CH and the malicious vehicle are $  \left\{Monitor, Not \ Monitor \right\}$ and $ =\left\{Attack, Wait\right\} $, respectively. The mixed-strategy NE of the game is obtained as $(p^*,q^*)$, where $p^*$ and $q^*$ are the probabilities that the malicious vehicle and the CH conduct $ Attack  $ and $ Monitor $, respectively.  By adopting the monitoring strategies based on the NE,  the volume of  IDS traffic is significantly reduced while maintaining a satisfied 
    detection rate and accuracy.

	    The cyber defense game \cite{brahmi2019cyber,sedjelmaci2018generic} for VNs takes the works in \cite{subba2018game} and \cite{sedjelmaci2014detection} a step further by using a hierarchical Stackelberg game. Not only does it consider the tradeoff between security and the overhead of IDS, but it integrates the functions of IDS, IPS, and Intrusion Reaction System (IRS) for detection, prediction, and reaction. This hierarchical game is played by an intrusion decision agent (IDA) (leader) and the secondary agents (i.e., IDS, IPS, and IRS). The IDA aims at balancing the communication quality and security by deciding the optimal activation of IDS, IPS, and IRS agents. An extensive-form representation of the game is given based on the payoff functions modeled for players.  By solving the game, the NE is obtained where the secondary players execute the corresponding strategies activated by the IDS. This scheme provides a new insight 
	    into the security game where IDS adaptively actives the actions of secondary players to detect, predict, or react against attacks with high accuracy, while considering the communication delay and overhead.

	    The DoS attack detection in VNs is addressed in \cite{mejri2016new} by modeling a zero-sum strategy game between the honest vehicle and the attacker. The attacker has two options: $ Attack  $or $ Stop $; the honest vehicle can $ Continue $ driving or $ Change direction $ to move away from the attacker. Misbehaving nodes identification in ephemeral wireless networks is discussed in \cite{behfarnia2019misbehavior} with a case study of VNs.  A voting-based Bayesian game between a benign node $PLB$ and a target node $ PLT $ is proposed. The strategies of $PLB$ and $PLT$ are $\left\{vote, abstain\right\}$ and $\left\{attack, not attack \right\}$, respectively. Moreover, incentives are used in payoffs in order to promote cooperation in the network. The pure-strategy and mixed-strategy BENs can help neighboring nodes decide whether to discredit an accused (target) vehicular node.

	\subsubsection{GT for privacy preserving}
	
		\begin{table*}
	\scriptsize
		\caption{Summary of Location Privacy Games in VNs}
		\label{tab_privacy}
		\renewcommand*{\arraystretch}{1}
		\begin{center}
			\begin{tabular}{|p{.02\textwidth}|p{.1\textwidth}|p{.24\textwidth}|p{.3\textwidth}|p{.05\textwidth}|p{.15\textwidth}|}
				\hline
				& \textbf{Game players}&\textbf{Location privacy metric}&\textbf{Description}&\textbf{Attack model}&\textbf{Description}\\
				\hline
				\cite{wang2019optimization}&$n$ vehicles in the mix zone&$P=\sum_{d=1}^{N} p_{d | b} \log _{2} p_{d | b}-\left(-\log _{2} \frac{1}{\Theta}\right)$&	\begin{itemize} [leftmargin=3pt] \setlength{\itemsep}{0pt}\item$ N $: the number of players\item$P_{b|d}$ the probability that the new pseudonym $ d $ of a specific vehicle is correctly linked by the attacker to its old pseudonym $b$\item$\Theta$ :eavesdropping station\end{itemize}&Passive attacker  &Eavesdrop \\
				\hline
				\cite{ying2015motivation}& $N_{\Upsilon}^{i}$ vehicles located in the mix zone &$B_{\Upsilon}^{i}=-\sum_{d=1}^{N_{\Upsilon}^{i}} P_{d | b} \log _{2} P_{d | b}$&\begin{itemize} [leftmargin=3pt] \setlength{\itemsep}{0pt}\item $N_{\Upsilon}^{i}$: the number of vehicles (located in the mix zone of $ Y $) during the $ i $ th round\item $Pd|b $: the probability that the new pseudonym $ d\in D $ of a specific vehicle is correctly linked by the attacker to its old pseudonym $ b \in B $\end{itemize} &Active attacker& Relate the target vehicle’s old pseudonym to its new pseudonym\\
				\hline
				\cite{plewtong2018game}&Defender and attacker&$L P_{i}(t)=L P M_{i}-L P L T_{i}(t)$ \newline $ L P M_i=\log _{2} n \forall n \geq 2 $\newline $L P L T_{i}(t)=x *\left(t-t_{s, i}\right)$&  \begin{itemize} [leftmargin=3pt] \setlength{\itemsep}{0pt}\item$ L P M_i$: the last mix-zone location privacy for vehicle $i$ \item$n$: the number of vehicles cooperate in the mix-zone\item$L P L T_{i}(t)$: location privacy loss over time\item $x$:  a constant for location privacy loss\item $ t_{s,i} $: last pseudonym change time\end{itemize}
				&\textit{Naive} attacker and  \textit{ Stealthy} attacker&\begin{itemize} [leftmargin=3pt] \setlength{\itemsep}{0pt}\item \textit{Naive}  attacker: never cooperates \item \textit{ Stealthy} attacker: attempts to obtain the location of a vehicle while not being detected\end{itemize}\\
				\hline
				\cite{fan2018network} & Legitimate node and malicious node&
				$ 
				\begin{aligned}
				&Privacy(\varphi, false, hit, dp) =\\&\sum_{r,r^{'}} \varphi(\mathrm{r}) \text { fake }<\mathrm{r} | \mathrm{r^{'}}>\mathrm{hit}<\widehat{\mathrm{r}}, \mathrm{r}>\mathrm{d}_{\mathrm{p}}(\widehat{\mathrm{r}}| \mathrm{r})
				\end{aligned}$
				&\begin{itemize} [leftmargin=3pt] \setlength{\itemsep}{0pt}\item$  \varphi (r) $: the probability that used by replacing the vehicle's real location information $  r  $ with other false location information \item $ fake <r,r^{'}> $: anonymity \item $ dp <\widehat{r},r> $: the premises of the know strategies of the players\end{itemize}
				& Active attacker&  Relate the target vehicle’s old pseudonym $ \widehat{\mathrm{r}} $ to its observed new pseudonym $ r' $\\
				\hline		
			\end{tabular}
		\end{center}
	\end{table*}

The privacy of vehicles should be carefully considered besides intrusion or malicious behaviors. The main concern for the privacy of a vehicle is the location privacy since an adversary can track the locations of vehicles by monitoring their pseudonyms. Anonymous authentication is a common scheme for location privacy preservation of vehicles in VNs, where vehicles temporally coordinate pseudonym changes using a mix-zone strategy to guarantee location privacy. To prevent the selfish motivations, several works \cite{ying2015motivation,plewtong2018game,plewtong2018game} use GT for stimulating cooperation among the vehicles in a mix-zone. Table \ref{tab_privacy} presents the comparisons of these location games.

	
    Wang et al. \cite{wang2019optimization} designed a non-cooperative game-based pseudonym changing scheme to balance the location privacy and the computation overhead in VNs. Each vehicle (player) in the mix-zone \cite{beresford2004mix} aims to maximize its location privacy with minimum computation overhead. The player performs the strategy $Cooperate$ (change its pseudonym) or $Defend$ (not change its pseudonym). The vehicle's utility is modeled as the location privacy according to the entropy of uncertainty and the eavesdropper's location (which is given in Table \ref{tab_privacy}). The NE strategy profile of the game is obtained corresponding to the optimal number of cooperating players $n*$. One weakness of this study is the use of a complete information game. In contrast, the incomplete information game could be better to model the knowledge of players because vehicles in the privacy-sensitive VNs do not know either the strategies or payoffs of their opponents.
	    
	In the non-cooperative games, vehicles only struggle to maximize their payoffs and will be unwilling to change the pseudonyms if they are satisfied with the current location privacy. It makes the target vehicle vulnerable to the attacks when more vehicles become selfish. Therefore, "vehicles with cooperative awareness" is the essential \cite{al2018teaming} for the location privacy in VNs. Ying et al. \cite{ying2015motivation} proposed a method called Motivation for Protecting Selfish Vehicles' Location Privacy (MPSVLP) to encourage vehicles to cooperate in changing their pseudonyms. In MPSVLP, vehicles form a mix zone dynamically when their pseudonyms are close to expiration. Moreover, an incentive mechanism is designed to encourage vehicles to implement pseudonym changes according to "reputation credit".

	In \cite{plewtong2018game}, a formal game model between the attacker and the defender is designed to analyze the impact of the attacker on location privacy in VNs. On the one hand, two Byzantine attackers that target location privacy in VNs are considered in \cite{plewtong2018game}: a \textit{naive} attacker who never cooperates and a \textit{ stealthy} attacker who attempts to obtain the location information of a vehicle while not being detected. On the other hand, the defender can cooperate to change its pseudonym because the location privacy could suffer a loss over time (the model is given in Table \ref{tab_privacy}); it can also be greedy and keep the old pseudonym due to the cost of pseudonym change.

	   Fan et al.  \cite{fan2018network} focused on the game between the legitimate node and the malicious node.The legitimate node replaces its real location information $r$ with a false location information $r'$  and sends $fake<r,r^{'}>  $ to the service provider. On the other hand, the malicious vehicle uses the received false location information as prior knowledge to form a rational conjecture $ hit<r,r^{'}> $ to predict the location of the attacked vehicle. The utilities of the legitimate player (i.e., location privacy metric) and adversary player are given in Table. \ref{tab_privacy}. However, the study only gives out maximization expressions of the players' payoff functions without NE solutions to the stated problem.

		A few recent studies use GT to prove the security of the privacy-preserving schemes in VNs. The security of the proposed authentication interchange scheme is proved by a valid argument using the strategies of attacker games in \cite{al2018teaming}. Studies in \cite{zhang2017otibaagka,zhang2019pa,mukherjee2019efficient} establish the security of their proposed privacy-preserving authentication schemes by defining a security game played between an adversary and a challenger in VNs.

	\subsubsection{ GT for trust management }
	  Trust management is designed to prevent false information transmitted by dishonest communication entities \cite{tian2019evaluating}, which is presented as a potential alternative of secure communication in VNs \cite{halabi2019trust}. Current trust evaluation models can be categorized as \cite{zhang2011survey}: 1) entity-oriented, which identifies the trustworthiness of the transmitter, usually based on interaction experience or neighbors’ recommendations; 2) data-oriented, which identifies the trustworthiness of transmitted messages; and 3) hybrid, which incorporates entity-centric and data-centric methods. 
	
	\begin{itemize}
\item\textbf{ Node-oriented approach}
	\end{itemize}
   Halabi and Zulkernine \cite{halabi2019trust} designed a distributed trust cooperation game to protect the data integrity of vehicles in IoV. Firstly, the vehicle's trustworthiness is constructed using the Bayesian inference model based on the previous interaction experience. Moreover, a hedonic coalitional game \cite{bogomolnaia2002stability} is designed for collaborations among vehicles according to their trustworthiness levels. The game is further solved by proposing a trust-based coalition formation algorithm. One advantage of this approach is that, since vehicles in the same coalition are not required to re-compute the trustworthiness values each time a new message is received, the computation overhead is reduced significantly.
	
		
	 In \cite{boudagdigue2018distributed}, a dynamic trust model for IoT is constructed where a set of neighbor vehicles and a monitored vehicle are considered as players.  In this model, the set of vehicles form groups dynamically assigning and updating the trust values of the monitored node according to the cooperativeness and honesty values.  An estimation algorithm is developed to remove spams caused by neighboring attackers during the trust value update. One weakness of this method is that the group is assumed to be formed in advance without considering the group formation process.

\begin{itemize}
	\item \textbf{Data-oriented approach}
\end{itemize}	

    Kumar \cite{kumar2015intelligent} proposed a Public Key Infrastructure (PKI) based on the Bayesian coalition game where the Learning Automata (LA) in the vehicles cooperate for information sharing. The dynamic coalition is formed based on the players' trust values formulated with symmetric key encryption and hash-based message authentication. The environment decides to reward or penalize the LAs according to their actions. Each player further updates the strategy vector, deciding whether to join a new coalition to improve communication security according to the feedback of the environment. The proposed scheme could resist against the replay and misauthentication attacks and preserve the message integrity and authentication.
	

 	Mehdi et al. \cite{mehdi2017game} proposed a game-based trust model to identify the defender nodes and the malicious nodes in VNs. The players can be tagged as a defender or malicious node according to the trust level calculated using the information provided by vehicles and RSUs (which is shown in Table. \ref{tab_trust}). Furthermore, the strategies of the defender and malicious node are defined as $  Detect /  not \ Detect  $ and $ Attack / not \ Attack $, respectively. And the payoff of each player for any combination of strategies is calculated. The NE is demonstrated where the defender can achieve satisfactory communication and avoid potential attacks. One limitation is that the trust model is based on a highway scenario without considering situations with high vehicle density, such as an urban environment.

		\begin{table*}
		\label{tab_trust}
		\caption{Summary of Trust Games in VNs}
		\scriptsize
		\renewcommand*{\arraystretch}{1}
		\begin{center}
			\begin{tabular}{|m{.04\textwidth}|m{.02\textwidth}|m{.1\textwidth}|m{.01\textwidth}|m{.01\textwidth}|m{.2\textwidth}|m{.27\textwidth}|m{.13\textwidth}|}
				\hline
				\textbf{Schemes}&\textbf{Ref.}&\textbf{Game}&	\rotatebox[origin=c]{90}{\textbf{Centralized}}&	\rotatebox[origin=c]{90}{\textbf{Distributed}}&\textbf{Trust metrics}&\textbf{Description}&\textbf{Attack resistance}\\
				\hline
				Node oriented&\cite{halabi2019trust}&Coalition game&&$\surd$&$Trust_{i}\left(V C_{m}\right)=\prod_{j \in V C_{m}} T_{i j}$\newline  \newline $Trust_i \ :\ V C_{m} \rightarrow[0,1]$ &Trust level of the coalition $ V Cm $ as perceived by vehicle $ V_i $.&Data alteration attack \newline Data corruption attack\\
				\cline{2-8}
					& \cite{boudagdigue2018distributed}&Assumed coalition game&&$\surd$&
					\begin{itemize} [leftmargin=3pt] \setlength{\itemsep}{0pt}\item Direct honesty: $p_{d}=D * p_{n e t}$ \item Indirect honesty:$p I=I * \text { pnet }$ \end{itemize} &
						\begin{itemize} [leftmargin=3pt] \setlength{\itemsep}{0pt}\item	$p_{D}$: the probability that the monitored node makes honest transactions\item$pnet$: the probability related to the constraints of the monitoring\item  $D$: direct honesty rate \item $p_l$: the probability that the monitored node evaluates honestly the other nodes\item$I$: Indirect honesty rate\end{itemize}&Spam\\
						\hline
						 Data-oriented&\cite{kumar2015intelligent} &Bayesian coalition game&&$\surd$&
						 \begin{itemize} [leftmargin=3pt] \setlength{\itemsep}{0pt}\item$\begin{aligned} T V^{t}&=\alpha \times  cert ^{rep }  PU ^{t}  (init) \\&-\beta \times cert ^{rev }P U^{t}  (curr) \end{aligned} $\item $ P U^{t}=\frac{cert^{curr} }{\sum_{i=0}^{n} cert _{i}^{tot}} $ \end{itemize}&\begin{itemize} [leftmargin=3pt] \setlength{\itemsep}{0pt}\item 	$ cert^{rev}(t) $: Certificate revocation at time $ t$  \item $cert_{i}(^{tot })$: total number of certificates \item $ P U^{t} (init)  $: Initial payoff utility \item  $ \alpha, \beta $:Constants with value between 0 and 1 \item  $ P U^{t} (curr)  $: Current payoff utility  \end{itemize}&Replay attack\newline Misauthentication attack\\
			 	\cline{2-8}
						& \cite{mehdi2017game}&Non-cooperative game&&$\surd$&$ L_{t}=\left(V_{e, t}-E_{T}\right) * 100 $ &\begin{itemize} [leftmargin=3pt] \setlength{\itemsep}{0pt}\item 	$L_t$: trust level of a node $ t $ \item $V_{e.t}$: event level raging from 0 to 1 experience based trust\end{itemize} &  Attacker prevention without focusing on specific attackers\\
				\hline
					Hybrid-oriented&\cite{fan2019trust} &Dynamic game&$\surd$&&$I B_{R S U}=\frac{n_{1}+n_{2}-n_{3}}{n_{1}+n_{2}+n_{3}}$&\begin{itemize} [leftmargin=3pt] \setlength{\itemsep}{0pt} \item $ IB_{RSU} $: recommendation value from RSU \item $n_1,n_2,n_3$: the number of forwarding, receiving, and releasing messages, respectively \end{itemize}&Restrain vehicles from selfish behaviors\\
				\cline{2-8}
					&	\cite{haddadou2014job} &Signaling game&&$\surd$&Distributed trust model (DTM2)& Allocate credits to nodes and securely manage these credits&Prevent both malicious and selfish vehicles\\
				\cline{2-8}
					&\cite{tian2019evaluating}&Evolutionary game&$\surd$&&$f _r \in\{x|1 \leq x \leq 100, x| \in Z\}$&$f_r$: the deception intensity of dishonest vehicles&Dynamic and varied attacking strategies\\
				\hline
			\end{tabular}
		\end{center}
	\end{table*}
	
\begin{itemize}
\item \textbf{ Hybrid approach}
\end{itemize}	

   To encourage nodes to forward data actively, Fan and  Wu \cite{fan2019trust} produced an incentive reputation mechanism based on a dynamic game. In this game, vehicle nodes are divided into three clusters of players: normal, selfish, and malicious nodes, each of which has its own strategy set. The RSU provides an incentive mechanism for the nodes within its range, i.e., rewarding the players who adopt cooperative behaviors and punishing those who are uncooperative. Vehicles in each cluster choose the strategies to maximize their payoffs by considering both the reward for cooperative behavior and the cost for energy consumption. The entire VN reaches an equilibrium after a definite number of evolution rounds of the repeated game. The proposed mechanism can restrain selfish behaviors and encourage vehicles to take cooperative actions.
	
	
    Facing the dynamic of vehicles and the openness of communication channels in VNs, trust relationships among vehicles should be constructed distributively without the reliance on a central entity. Haddadou \cite{haddadou2014job} proposed an infrastructure-independent trust model $ (DTM^2)  $ to prevent both malicious and selfish behaviors in VNs. The $ (DTM^2)  $ maintains 
     \textit{credits} and allocates them to each vehicle.  Firstly, selfish nodes are motivated to cooperate for credits earning by establishing the price for receiving messages. Furthermore, to detect the malicious nodes, $  DTM^2  $ constructs the cost of sending by formulating signal values corresponding to the behaviors of the nodes. Nodes whose messages are identified trustful will be rewarded with \textit{credits}.  Lastly, the credit update process of a vehicle is modeled as a Markov chain based on the vehicle's behavior.
	
    Most of the credit management schemes mainly focus on certain types of attacks with settled attacking behaviors, which is insufficient to represent the real scenarios.  In \cite{tian2019evaluating}, a dynamic and varied attacking strategy is designed to evaluate the reputation of communication entities and trustworthiness of messages. An evolutionary game is employed to analyze the evolution of attackers' strategies, where untrustworthy vehicles will be removed from the network. The method enables  simultaneous detection the credibility of the received messages and the trustworthiness of the communication vehicles.
	
	
	\subsection{Game Theory for QoS Guarantee in VNs}

Exploiting available resources to maximize the overall QoS of VNs, such as throughput improvement, is another challenge in VNs. GT is capable of analyzing the behaviors of nodes to help them decide the optimal strategies and further optimize the performance of the network. The application of GT in QoS guarantee for VNs has been studied in various ways, such as designing cooperative routing, MAC protocols, spectrum resource allocation, and power control.

	\subsubsection{GT for routing}
\begin{figure*}[!hbt] 
	\centering
	\includegraphics[width=7.3in]{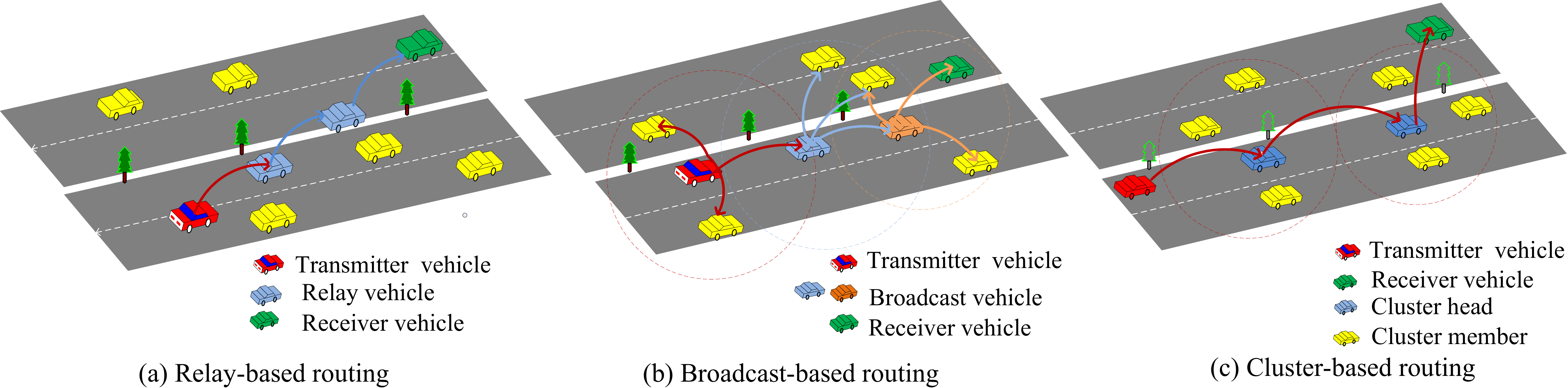}
	\caption{Routing schemes in VNs: (a) relay-based routing,(b) broadcast routing, (c) cluster-based routing }
	\label{fig_routing}
\end{figure*}

GT has been widely applied for mitigating selfish behaviors of packet forwarding in wireless networks. However, maintaining stable routes in VNs is challenging due to the dynamic nature of the mobile vehicle. Routing schemes in VNs can be categorized as relay-based, broadcast-based, and cluster-based routings, which is illustrated in Fig. \ref{fig_routing}.

	\begin{itemize}
		\item \textbf{Relay-based routing}
	\end{itemize}
Tian et al. \cite{tian2017self} proposed a non-cooperative game-theoretic method where the candidate relay nodes are treated as players who aim to optimize the transmission quality with low energy consumption. The utility function $u_j$ of each player $j$ is formulated by considering the trade-off between the received benefit and the energy or power cost, which is given as (\ref{eq_Utility_tian2017self}). According to the game structure, a decentralized self-organized algorithm is further proposed to adapt the decision-making behavior of each player and learn their NE strategies from their action-reward histories. The NE can preserve the transmission reliability and maintains the overall power consumption at the lower bound.

	\begin{equation}
	\label{eq_Utility_tian2017self}
	u_{j}\left(\boldsymbol{s}_{j}(t)\right)=f_{j}\left(\boldsymbol{s}_{j}(t)\right)-g_{j}\left(\boldsymbol{s}_{j}(t)\right), 
	\end{equation}
	where $f_j$ and $g_j$ denote the obtained benefit and the cost incurred by the cooperation of the relay node $j$ when adopting the strategy $s_j (t)$ over time $t$, respectively.


   To mitigate transmission failure, the authors in \cite{das2017new} proposed a game-theoretic relay selection protocol to select the cooperative node in the retransmission. This algorithm stimulates nodes to cooperate by providing incentives and selects the best relay nodes for retransmission of packets when the communication link becomes unstable. The network can converge into a Nash network in the highly dynamic VN environment. This protocol shows lower transmission delays compared to non-cooperative communications. 
	
	  Suleiman et al. \cite{suleiman2017adaptive} employed the GT to demonstrate the incentive-compatibility of the routing protocol for highway applications in VNs. A fixed cellular network operator $ CO $ and a set of vehicles $ V $ constitute the players of the game in the highway scenario. On the one hand,  $ CO  $ chooses Incentivizing Cooperation $  (IC) $ making credit-based bandwidth allocation or Not Incentivizing Cooperation $ (NIC) $ fixing bandwidth allocation. On the other hand, the strategy of any vehicle $  (AV_i \in V )  $ is Cooperation $ (C) $ relaying other vehicles’ messages or Not Cooperating $ (NC) $ dropping the messages. Satisfied routing strategy can be obtained by NE, where the incentive V2V routing complements V2I routing. However, this method is infrastructure-based and only designed for highway applications, which can not be applied to the infrastructure-independent communications or the city scenarios.
	

	In order to overcome the limitation of traditional Nash Bargaining Solution (NBS) that lacks the complete information,  Kim \cite{kim2016timed} proposed a new opportunistic routing scheme for dynamic VNs based on an iterative bargaining model. By considering the timed game, the destination vehicle selects the most adaptable routing path $\Gamma$ with the minimum link cost:
	
	\begin{equation}
	\label{eq_Path_kim2016timed}
	\Gamma=arg \underset{\{P_i, P_i\in \mathbb{S}\}}{min} {\int_{t=\mathcal{T}_s}^{t=\mathcal {T}_{e}} \log \left(L_{-} P_{k}\right) d t}
	\end{equation}
	    where $ \mathbb{S} $ is the set of established routing paths, and $  P_i $ is the $ i $th routing path from the source vehicle to the destination vehicle. $\mathcal{T}_s$ and $\mathcal{T}_e$ are the packet forwarding start time and end time, respectively, and  $L_P$ denotes the communication cost of each link. The destination vehicle periodically updates the timed reinforcement function to stimulate relay vehicles to select desirable links globally.

	\begin{table*}
		\footnotesize
		\label{tab_routing}
		\caption{Summary of routing games in VNs}
		\renewcommand*{\arraystretch}{.5}
		\begin{center}
			\begin{tabular}{|m{.02\textwidth}|m{.08\textwidth}|m{.1\textwidth}|m{.15\textwidth}|m{.13\textwidth}|m{.09\textwidth}|m{.06\textwidth}|m{.02\textwidth}|m{.01\textwidth}|m{.01\textwidth}|m{.01\textwidth}|m{.01\textwidth}|}
					\hline
				\textbf{Ref.}&\textbf{Game}&\textbf{Players}&\textbf{Objective}&\textbf{Main Idea} &\textbf{Routing}&\textbf{\textbf{Scenario}}& \rotatebox[origin=c]{90}{\textbf{Speed}}&\rotatebox[origin=c]{90}{\textbf{Position}}&\rotatebox[origin=c]{90}{\textbf{Inter-distance}}&\rotatebox[origin=c]{90}{\textbf{Acceleration}}&
					\rotatebox[origin=c]{90}{\textbf{Channel Model}}\\
					\hline
					\multicolumn{12}{|c|}{\multirow{2}{4cm}{\textbf{Relay-based Routing}}}\\
					\multicolumn{12}{|c|}{}\\
					\hline
					\cite{tian2017self} &Non-cooperative game &Candidate relay nodes&Optimize transmission quality with low energy consumption&Incentive best relay selection&Centralized&Urban&$\surd$&$\surd$&$\surd$&$\surd$&$\surd$\\	
					\hline			
				     \cite{das2017new}&Non-cooperative game&Neighbor nodes&Retransmission in case of network failure&Best relay selection&Centralized&$-$&&&&&\\
					\hline
					\cite{suleiman2017adaptive} &"One shot game"&A fixed cellular network and vehicles&Credit-based bandwidth allocation&Best routing selection among V2I and/or V2V techniques&Hybrid&High-way&$\surd$&$\surd$&$\surd$&$\surd$&$\surd$\\
					\hline
					\cite{kim2016timed}&Bargaining game& Vehicles&Opportunistic routing selection&Best routing selection&Decentralized&$-$&$\surd$&&$\surd$&&\\
					\hline
						\multicolumn{12}{|c|}{\multirow{2}{6cm}{\textbf{Broadcast-based Routing}}}\\
						\multicolumn{12}{|c|}{}\\
					\hline
					 \cite{hu2016novel}&Non-cooperative game&Vehicles&Delay minimization for emergency message broadcast & Rebroadcast&Decentralized&$-$ &$\surd$&&$\surd$&&$\surd$ \\
					 \hline
					  \cite{assia2019game}  &Non-cooperative game&  Vehicles& Reachability maximization and delay minimization &Rebroadcast&Decentralized&Grid-map&$\surd$&$\surd$&$\surd$&&\\
					\hline
						\multicolumn{12}{|c|}{\multirow{2}{6cm}{\textbf{Cluster-based Routing}}}\\
						\multicolumn{12}{|c|}{}\\
					\hline
					   \cite{sulistyo2019coalitional}&Coalitional game&Vehicles &Cluster stability maintaining and link quality improvement&Optimum cluster formation and CH selection&Decentralized& City toll&$\surd$&&&&$\surd$\\
					   \hline
					    \cite{huo2016coalition} &Coalitional game & Vehicles&Trade-off between stability and efficiency of communication & Optimum cluster formation and CH selection &Decentralized&Crossroad&$\surd$&$\surd$&&&\\
					    \hline 
					    \cite{dua2017reidd}&Coalitional game &Vehicles&Broadcast storm mitigation&Optimal cluster formation and CH selection&Hybrid&Patiala city&$\surd$&&$\surd$&$\surd$&\\
					    \hline
					    \cite{khan2018evolutionary} &Evolutionary game&Clusters and vehicles& Throughput optimization for cluster\newline Routing stability maintaining &Optimum cluster formation and CH selection&Hybrid&Manhattan grid&$\surd$&&&&\\
					    \hline
					    \cite{wu2018computational} &Coalitional game&One-hop neighbors of RSU or CH &Multi-hop routing optimization&Dynamic cluster formation and CH selection&Hybrid&Freeway and street&$\surd$&&$\surd$&&$\surd$\\
					    \hline					    
				\end{tabular}
			\end{center}
\end{table*}


	\begin{itemize}
		\item \textbf{Broadcast-based routing}
	\end{itemize}

In VNs, the broadcast storm may happen when the network is overwhelmed by continuous broadcast or rebroadcast traffic. The simplest scheme of the broadcast is blind flooding \cite{tseng2002broadcast}, where each vehicle node rebroadcasts the received messages until all vehicles in the network receive the messages. However, in a high-dense network, flooding will cause a large amount of redundant information, which obviously leads to collision and congestion. On the other hand, in the environment with sparse traffic, a single forwarder can not maintain the connection and reachability of the communication. Therefore,  cooperative forwarding is essential for vehicle nodes to cover the network with a low frequency of rebroadcast.

	Emergency message warning is highly time-critical and requires a more intelligent broadcast mechanism with low latency and high reliability. In \cite{hu2016novel}, an emergency message broadcast game is proposed that takes vehicle density and link quality into account. In this scheme, each vehicle calculates the link quality using the receiving power. Then, the probability of rebroadcasting the emergency message is determined based on the NE. The proposed rebroadcast scheme obtains low broadcast overhead and transmission delay compared to blinding flooding.
	
    To maximize reachability while minimizing delay and the number of rebroadcasts of VNs,  Assia et al. \cite{assia2019game} designed an effective rebroadcasting protocol. The authors presented an adaptive algorithm by modeling cooperative forwarding as a volunteer’s dilemma game. Based on the game, each vehicle uses the received information such as speed and position to decide whether to rebroadcast. Simulation results show that this protocol outperforms prior work on the reachability, the number of rebroadcasts, and the delay, especially in congested areas.
	
	
	\begin{itemize}
	\item \textbf{{Cluster-based routing}}
	\end{itemize}
	
	Clustering can be considered as an efficient solution to enhance the connectivity of VNs.  The coalitional game is considered as a potential approach for clustering formation. It has been used in mobile ad hoc networks (MANETs) \cite{massin2017coalition} because the concept of the cluster is compatible with the coalition. However, the limited communication resources and high mobility of vehicles make it difficult for vehicles to maintain a balance between the cluster stability and communication efficiency. The clustering protocols for VNs can be categorized as \textit{centralized clustering} based on a central controller (e.g., roadside unit), \textit{decentralized clustering} based on V2V communication, or \textit{ hybrid clustering} which incorporates the above two protocols.  However, centralized clustering is not flexible for VNs because it is dependent on the roadside infrastructure. Therefore, most of the recent GT-based clustering schemes for VNs aim to balance the cluster stability and clustering efficiency using decentralized  \cite{sulistyo2019coalitional,huo2016coalition} or hybrid clustering  \cite{dua2017reidd,khan2018evolutionary,wu2018computational}.
 
 	Sulistyo et al. \cite{sulistyo2019coalitional} proposed a stable clustering method for V2V communication based on the coalitional game to improve the channel capacity. In the game, each vehicle aims to form a cluster by establishing V2V links with its neighbors based on the coalition values. The coalition value is formulated as the function of SNR, link lifetime, and velocity difference among vehicles, all of which quantifies the V2V communication quality.  Based on the coalition value, vehicles can decide whether to establish a new connection so that to obtain a good trade-off between cluster stability and communication quality. 
 	
 	Focusing on crossroad scenarios,  Huo et al. \cite{huo2016coalition} presented a coalition game-based clustering strategy considering both efficiencies of communication and stability of the cluster. The vehicles make decisions whether to switch to a new coalition according to the coalition utility, which is formulated by the relative velocity, relative position, and the bandwidth availability ratio of vehicles. The coalition formation can converge to a Nash-stable partition, in which case the bandwidth of the CH can be fully utilized, and the stability of the cluster can be guaranteed.


To mitigate the broadcast storm,  Dua et al. \cite{dua2017reidd} proposed a game-based reliable data dissemination protocol in VNs. Firstly, vehicles (players) communicate with their neighbors to calculate their payoffs by estimating the stability of the links. The players dynamically form clusters based on the regularly updated payoffs, and the CH informs the RSU about the vehicles in the cluster. For data dissemination, the scheme can determine the best route from the source to destination by selecting the link with maximum payoff value rather than blindly flooding. In the case where the route from the source and destination is not connected, the nearest RSU will join this game using the store-carry-forward technique.

	 Khan et al. \cite{khan2018evolutionary} proposed an evolutionary game-based scheme to improve the throughput and stability of clusters. This approach employs a semi-distributed clustering method. On the one hand,  the communication throughput of all clusters is managed centrally by the RSU. On the other hand, the cluster formation and CHs nomination are automated in a distributed way. The existence of the equilibrium point is demonstrated analytically, and its stability is proved with the Lyapunov function.

To improve the efficiency of channel contention at the MAC layer, Wu et al. \cite{wu2018computational} designed a multi-hop routing protocol for (Vehicle-to-Roadside-unit) V2R communication based on a hybrid clustering. In this scheme, a coalitional game is employed to stimulate the vehicles to join a cluster, in which the CHs act as relays for multi-hop data transmission. An RSU or CH acts as payoff allocator and distributes payoffs to its one-hop neighbors who constitute the players of the game. Therefore, each vehicle would like to cooperate for data forwarding because only RSU or CH assigns the payoff. Besides, a fuzzy logic algorithm is designed for dynamic CH selection by using multiple metrics, such as speed, signal qualities, and moving patterns of vehicles. By selecting CH for data forwarding, the number of transmitter can be reduced, and the channel access contention can be relieved consequently.

		\subsubsection{GT for MAC protocols design}
	Similar to traditional wireless networks, the design of the MAC layer protocols for VNs is challenging due to vehicles' competition for accessing the limited bandwidth. Furthermore, some vehicle nodes may deviate from the MAC protocol to obtain an unfair bandwidth share. To solve these problems, GT has been applied by several studies with the purpose of collision alleviation \cite{kwon2016bayesian,li2019tcgmac}, competition solution \cite{wang2018application,lang2019vehicle}, and greedy behaviors prevention \cite{al2017cooperative}.
	    
	  Kwon et al.  \cite{kwon2016bayesian} proposed a Bayesian game-theoretic approach for the beacon collision alleviation at the MAC layer in urban VNs. Each selfish or cooperative vehicle makes individual decisions regarding the power level or transmission probability based on its payoff. The payoff of each vehicle is measured in terms of throughput rewards and transmission energy costs. The authors proved the existence of the BNE of the game, where vehicles can achieve higher utilities and fairness compared with the random access. Collision alleviation is also studied in \cite{li2019tcgmac} by designing a MAC protocol that synthesizes Time Division Multiple Access (TDMA) and Carrier Sense Multiple Access (CSMA) mechanisms. This scheme divides the time frame into two segments: the TDMA period used for application data transmission and CSMA period used for slot declaration. GT is employed to maximize the usage of slots by deferring the nodes' choices when two or more nodes reserve the same slot during the CSMA period.

    A medium access algorithm is designed in \cite{wang2018application} to solve the competition for access to the wireless channel of VN. The concept \textit{application value awareness} is proposed by assigning different values to packets according to the waiting time during the competition so that the nodes with higher packet values have higher channel access probability. A cross-layer game model is then constructed for the players (vehicles) to adjust their channel access probabilities (strategies) based on their packet values. The equilibrium enables the successful transmission of the message within the latency limit when the channel is nearly saturated. The inter-distance between vehicles and the roles of vehicles are further considered in the medium access game in \cite{lang2019vehicle} to determine a player's priority to access the channel.
	
    Greedy behavior prevention in VNs with Carrier Sense Multiple Access with Collision Avoidance (CSMA/CA) MAC protocol is considered in \cite{al2017cooperative} using a game mechanism that encourages the selfish nodes to behave normally under the threat of retaliation. The authors proposed group reputation based and cooperative detection based tit-for-tat strategies to impose cooperation among vehicles. This adaptive CSMA/CA protocol can motivate the selfish nodes to cooperate under the threat of retaliation and is immune to the ambiguous monitoring caused by collisions.

	\subsubsection{GT for power control}
	\begin{table*}
	\scriptsize
	\label{tab_powercontrol}
	\caption{ Summary of power control games in VNs}
	\renewcommand*{\arraystretch}{1}
	\begin{center}
		\begin{tabular}{|m{.02\textwidth}|m{.1\textwidth}|m{.13\textwidth}|m{.13\textwidth}|m{.17\textwidth}|m{.15\textwidth}|m{.01\textwidth}|m{.01\textwidth}m{.01\textwidth}m{.01\textwidth}|}
			\hline
			\textbf{Ref.}&\textbf{Game}&\textbf{Objective}&\textbf{Strategy}&\textbf{Utility Metric}&\textbf{Solutions}&\rotatebox[origin=c]{90}{\textbf{Fairness}}&\rotatebox[origin=c]{90}{\textbf{Distributed}}&\rotatebox[origin=c]{90}{\textbf{Centralized}}&\rotatebox[origin=c]{90}{\textbf{Hybrid}}\\
			\hline
			\cite{goudarzi2018non}& Non-cooperative game&Beaconing congestion control&Beacon transmission power control&Channel load, channel busy ratio (CBR) and price function&Theoretic NE calculation&$\surd$&$\surd$&&\\
			\hline
			\cite{goudarzi2019fair}  &Non-cooperative game&Beaconing congestion control &Joint beacon frequency and power control&Channel load, CBR, and price function&NE and gradient dynamics&$\surd$&$\surd$&&\\
			\hline
			\cite{shah2018shapely}& Cooperative game& Beaconing congestion control &Adaptive transmit power control&Marginal contributions of vehicles towards congestion&Shapley value model &$\surd$&$\surd$&&\\
			\hline
			\cite{amer2017game} &Non-cooperative game &Data channel congestion alleviation&Dynamic data rate adaptation &Data rate and priority of each vehicle&Karush-Kuhn-Tucker conditions and Lagrange multipliers&$\surd$&&&$\surd$\\
			\hline
			\cite{hua2017game}&Non-cooperative game& Balance between energy consumption and QoS &Adaptive transmit power control&Throughput, delay, and channel model&NE,Pareto Optimality and Social Optimality&&$\surd$&&\\
			\hline
			\cite{chen2016information}&Bargain game&Information congestion control for intersection scenarios&Joint power and packet generation rate control&Channel load&Clustering and CH selection&&$\surd$&&\\
			\hline
			\cite{sun2017non}&Non-cooperative game&Balance between QoS and security&Transmit power and the encryption block length control&Channel capacity&Theoretic NE calculation&&$\surd$&&\\
			\hline
		\end{tabular}
	\end{center}
\end{table*}

Power and rate control is essential to avoid congestion caused by the uncertainty of communications in VNs such as unstable channel, dynamic vehicles, and varied environments. In VNs, vehicles periodically broadcast their status in Basic Safety Messages (BSMs), also known as beacons, to make the neighbors aware of their presence. Higher power or rate is desired by a vehicle to disseminate the messages over a larger distance. However, a high level of power could cause collision or congestion to communication in dense traffic environments, leading to both performance and security degradation. Therefore, a good power or rate control scheme should adapt to varying environments, which should provide sufficient awareness of surrounding vehicles’ status while maintaining low congestion and high security of communication.  GT may contribute to the prediction of vehicles’ behaviors, and could guide rational choices of vehicles’ power or rate according to the environment in VNs.
 

In \cite{goudarzi2018non}, the congestion control is designed from the perspective of adaptive BSM power adjustment.The algorithm employs a non-cooperative game $ \mathcal{G}=\left\{\mathcal{N},\left\{\mathcal{P}_{i}\right\}_{i \in \mathcal{N}},\left\{\mathcal{F}_{i}\right\}_{i \in \mathcal{N}}\right\} $, where $\mathcal{N}=\{1,2, \ldots, N\}$ is the player set, $ P_i  $ is the set of possible beaconing powers of player $  i  $, and $ F_i $ is the payoff function.  The payoff function of player $i$ is presented as:

\begin{equation}
	\label{eq_payoff_goudarzi2018non}
	\begin{aligned}
		\mathcal{F}_{i}\left(p_{i}, \mathbf{p}_{-i}\right) &=U_{i}\left(p_{i}\right)-J_{i}\left(p_{i}, \mathbf{p}_{-i}\right) \\ &=u_{i} \ln \left(p_{i}+1\right)-\frac{c_{i} p_{i}}{1-C B R_{i}\left(p_{i}, \mathbf{p}_{-i}\right)}
	\end{aligned}
\end{equation}
where $  u_i $ and $ c_i $ are positive parameters, and $ C B R_{i}\left(p_{i}, \mathbf{p}_{-i}\right )$ denotes the Channel Busy Ratio (CBR) sensed by player $  i $, which is a function of all players’ beacon power. Based on the game model, the existence and uniqueness of the NE are demonstrated theoretically. Furthermore, the scheme attains  stability, fairness, fast convergence, and efficiency in controlling the congestion below target levels.

Vehicles would reduce both beacon power and frequency in situations with dense traffic. Although Decentralized Congestion Control (DCC) is a joint beacon rate and power control mechanism \cite{etsi102decentralized}, it suffers unfairness and oscillation \cite{rostami2016stability}. Such problems can be avoided in \cite{goudarzi2019fair}, where a joint beacon rate and power control game is developed to prevent congestion in the shared channel. The NE is obtained using gradient dynamics; the uniqueness and stability of the NE are proved mathematically. Simulation results indicate that the fairness of beacon power control can be achieved without causing extra overheads, and CBR can be controlled at an appropriate level.

    In \cite{shah2018shapely}, an adaptive transmit power control for vehicular communication is developed for vehicles to adapt the transmission powers to local channel congestion. This approach employs a cooperative game where players (vehicles) cooperate and join a coalition during a congestion event. Fair power adaptation of a vehicle is obtained using the marginal contributions of the Shapely value model. This power control scheme could effectively transfer marginal contributions into a fair power decrease to avoid congestion

	Amer et al. \cite{amer2017game} proposed a dynamic data rate adaptation approach based on a non-cooperative game. Each vehicle is modeled as a selfish player in the game who requests a high data rate as its strategy. The utility function of each vehicle is formulated according to vehicles’ desires for high data rates (payoff function) and the priority of the vehicle to achieve satisfied fairness (priority cost function). The optimal game data rate is obtained by using Karush-Kuhn-Tucker conditions and Lagrange multipliers, which could satisfy congestion mitigation and provide a fair allocation for network resources.
	
	Chen et al.  \cite{chen2016information} focused on the information congestion control for intersection scenarios of VNs. The authors designed a joint power and packet rate adaptation method based on a bargaining game.  Each CH acts as a player in the game and determines the optimal transmission power and packet generation rate for its cluster members. The utility of each player is constructed using channel load monitoring and estimation. Simulation results demonstrate that the scheme outperforms IEEE802.11p concerning queuing delay and throughput.

To guarantee the green communication in VNs, Hua et al. \cite{hua2017game} proposed a repeated power efficiency game model considering the throughput and delay performances of VNs. In the game, each vehicle performs a packet forwarding strategy, i.e., power efficiency. Furthermore, NE is presented, and its efficiency is analyzed using Pareto and social optimality. Based on NE, each vehicle dynamically adjusts its transmission power to make a balance between energy consumption and throughput. Simulation results reveal that the power efficiency game can provide efficient strategies in green communication.  However, the method does not consider the fairness problem.

Sun et al. \cite{sun2017non} proposed an adaptive transmit power and encryption block length control to balance the QoS and security by allocating the limited computing resources in VNs. This scheme is based on a non-cooperative game with a “communication player” who controls the transmit power and a “security player” who decides the encryption block length. By calculating NE value, the optimal transmit power and encryption block length are obtained. However, the game model is simple because it only focuses on a single-vehicle without considering the competition or cooperation behaviors among multiple vehicles.

	\subsubsection{GT for spectrum resources allocation}
   Fair and effective spectrum resources allocation has become one of the primary tasks in VNs due to the selfishness of vehicles and the scarcity of spectrum resources. Cognitive Radio (CR) \cite{akyildiz2006next,bayhan2013overview} is a spectrum sharing concept which allows unlicensed vehicles to access licensed spectrum in a non-intrusive manner and is envisaged as a potential technology to deal with the spectrum scarcity in VNs. The network formed by those nodes is called CR-VNs, which aims to enhance the connectivity of vehicles and improve the utilization of constrained spectrum resources \cite{he2016resource}.

Optimal network selection is investigated in \cite{kumar2016spectrum}, where a spectrum handoff scheme based on auction game in CR-VNs is proposed.  The authors used the decision-making method with multiple attributes to formulate the cost functions of players. The proposed scheme adapts to different types of CR vehicular nodes and multiple available networks. However, how to extend this scheme to dynamic environments is not taken into consideration.

In  \cite{shattal2018channel}, the dynamic channel selection strategy is studied with the purpose of joint optimizing the infotainment throughput and spectrum access in VNs. This approach views the channel selection as an evolutionary game where vehicles act as players who continuously select a strategy. Three models of strategy are considered in this work: 1) always consume user who chooses the channel randomly, 2) forage-consume user  who senses the channel using the collected information, and 3) social-forage consume user who sometimes waits and allows other users access exclusively. The RSU firstly collects information of vehicles within its range and relays it to the cloud. Then the cloud computes the optimal and stable strategy using the evolutionary game and sends the results to RSU. At last, the RSU broadcasts the recommended strategies to the vehicles within its radio range. The experiment results show that a significant improvement in the network performance can be obtained when vehicles employ  evolutionary strategies recommended by RSU.

Joint allocation of subcarrier and transmit power \cite{eze2019design} is proposed based on NBS to support the enhanced packet transmission in CR-IoVs. The cognitive vehicular Secondary Users (SUs) form clusters to opportunistically access the shared channels, which are allocated to the Primary Users (PUs) on the condition that the Signal-to-Interference-plus-Noise-Ratio (SINR) value of the PUs is below the threshold level. Furthermore, the optimal subcarrier and transmit power allocation strategies are derived based on the NBS. Besides, an algorithm with iteration-independence and low-complexity also ensures the convergence to Pareto optimality.
	
    Considering the inefficiency of the NE, Tian et al. \cite{tian2019channel} proposed a channel pricing mechanism using marginal social cost to motivate selfish vehicular nodes to achieve a social optimum of the spectrum allocation. The authors considered a CR-enabled V2I communication scenario where the cooperation between exclusive-used and shared-used spectrum resources is modeled as an evolutionary game. They further proved that the NE of the proposed game is evolutionarily stable and coincides with the social optimum.

	\section{Game Theory in Next Generation VNs}
	\label{sec_GTin5G}

 Next 5G communication technology provides the potential to meet the demands of a large number of connected IoT devices and various data-intensive applications. 5G provides low-latency, high-reliability, ubiquitous, and energy-efficient communications \cite{xiang20165g}. With the rigorous requirements of 5G, emerging technologies such as edge computing are integrated into VNs to promote the evolution of current VNs towards 5G-based VNs. In what follows, the recent contributions of GT to diverse emerging technologies integrating with VNs such as Mobile Edge Computing (MEC), Heterogeneous Vehicular Networks (HetVNETs), Software Defined Networking (SDN), and Unmanned Assisted Vehicles (UAV) are presented.

\subsection{Game Theory in Vehicular Edge Computing}
Vehicular Edge Computing (VEC) is an emergent architecture that offers cloud computing capabilities at the edge of the VNs to manage the computation-intensive and real-time tasks with short latency. However, the problems of task offloading, content delivery, and resource allocation or sharing are challenging in VEC due to the limited storage capacity of Mobile Edge Computing (MEC), especially in dense traffic environments where vehicles have quantity service demands. Furthermore, the lack of complete information between the tasks and edges makes the problem more complicated. GT, which is a powerful tool to deal with conflicts, can effectively address the decision-making concerns among multiple users who compete for limited resources. 

\subsubsection{Overview of VEC}
\begin{figure*}[!hbt] 
		\centering
	    \includegraphics[width =6.6in]{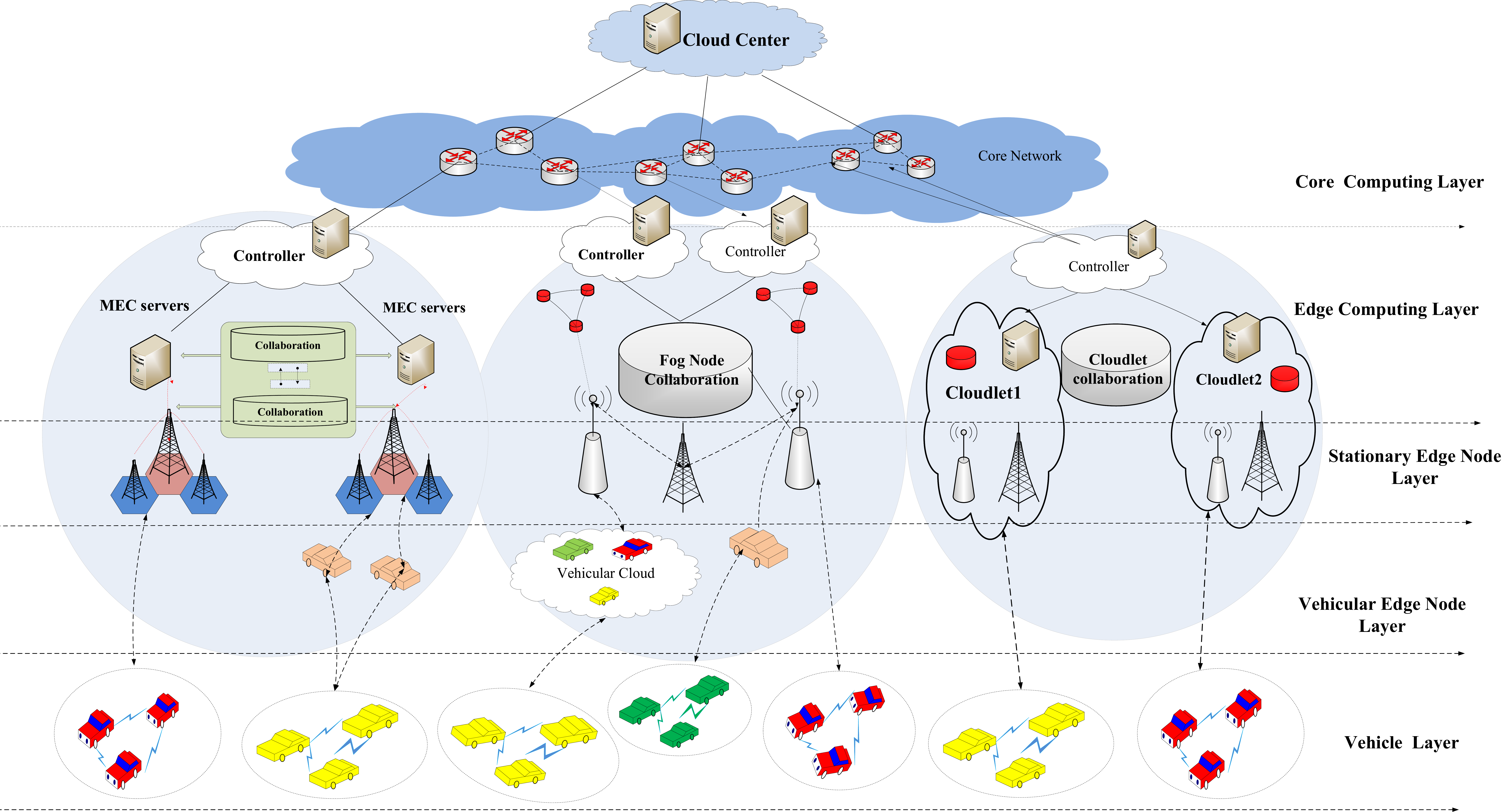} 
	    \caption{The architecture of VEC (primarily based on \cite{wang2018enabling}) } 
	     \label{fig_vec}  		
	\end{figure*}

As is shown in Fig. \ref{fig_vec}, the three-layer VEC architecture consists of the cloud layer, edge computing layer, and vehicular layer. At the edge computing layer, the implementations for VEC can be classified into MEC, fog computing, and cloudlet computing  \cite{dolui2017comparison}. At the vehicular layer, Vehicular Cloud (VC), as an extension of edge layer architectures, plays an essential role in assisting the edge layer or cloud layer with effective resource management.

From the perspective of the participating nodes, edge nodes are commonly characterized into stationary edge nodes and vehicular edge nodes \cite{dziyauddin2019computation}, which are at the stationary edge node layer and vehicular edge node layer, respectively.

\begin{itemize}[leftmargin=15pt]\setlength{\itemsep}{10pt}
	\item \textbf{Stationary edge nodes} An roadside infrastructure such as RSU and BS connected to a MEC server or a fog server can serve as stationary nodes.  Besides, a cloudlet consisting of a set of roadside infrastructures also plays the role of an edge node that provides computation and storage capabilities to vehicles.
	
	\item \textbf{Vehicular edge nodes} Vehicles with available computation resources and capabilities can form cloudlets or VCs for resource sharing.
	
	\item \textbf{Hybrid edge nodes} Stationary and vehicular edge nodes coexist in the VEC.
	
\end{itemize}Table. \ref{tab_edgenode} presents the types of edge nodes in VEC. The applications of GT in VEC are discussed from the aspects of MEC, FC, cloudlet, and VC receptively in the following subsections.

	\begin{table*}
		\footnotesize
		\caption{VEC edge nodes}
		\label{tab_edgenode}
		\renewcommand*{\arraystretch}{1}
		\begin{center}
			\begin{tabular}{|m{.16\textwidth}|m{.26\textwidth}|m{.28\textwidth}|}
				\hline
				\textbf{Categories} &\textbf{Types of VEC nodes}&\textbf{Related works}\\
				\hline
				Stationary Edge Nodes
				&Roadside infrastructure with MEC server & \cite{zhao2019computation},\cite{liu2018computaion}, \cite{zhang2017mobile},\cite{zhang2019task},\cite{zhang2017optimal}, \cite{hui2019edge},\cite{zhou2018begin},\cite{huang2017distribute}  \\
				\cline{2-3}
				&Roadside infrastructure with fog server& \cite{kang2017privacy},\cite{zhang2017resource}\\
				\cline{2-3}
				&Roadside infrastructure cloudlet&\cite{yu2016optimal}, \cite{tao2017resource} \\
				\hline
				Vehicular Edge Nodes&Vehicles& \cite{xu2018low},\cite{klaimi2018theoretical},\cite{aloqaily2017fairness},\cite{brik2018gss}, \cite{mekki2017proactive},\cite{liwang2019game}\\
				\cline{2-3}
				&SP cloudlet &		 \cite{yu2015cooperative}, \cite{lin2019vehicle}\\
				\hline
				Hybrid Edge nodes&MEC server and vehicles&\cite{gu2019task}\\
				\hline
			\end{tabular}
		\end{center}
	\end{table*}

\subsubsection{MEC-implemented VEC }
	Several recent studies use GT to deal with the optimal decision-making on computation or task offloading \cite{liu2018computaion,zhang2017mobile,zhang2019task,gu2019task}  to reduce the computation delay and overheads. Furthermore, some works also consider resource allocation by jointly optimizing the task offloading for vehicles and the resource assignment for service providers (SPs) \cite{zhao2019computation,zhang2017optimal,huang2017distribute,zhou2018begin}. Besides, content delivery with GT solution in MEC-implemented VEC is addressed in \cite{hui2019edge}. Table. \ref{tab_MEC} summarizes GT-based schemes in MEC-implemented VEC.

	\begin{itemize}
	 \item  \textbf{Computation or task offloading  game}
	\end{itemize}

    Liu et al. \cite{liu2018computaion} modeled the decision-making of computation offloading among vehicles as a multi-user potential game, where each vehicle chooses to use the local computing or cloud computing to minimize its computation overhead.  The existence of NE is proved, and the performance of the computation offloading is quantified by the anarchy price.

A cloud-based MEC offloading framework in VNs is proposed in \cite{zhang2017mobile} to reduce the transmission cost. This network scenario consists of $M$ RSUs with the same coverage $L$, each of which has a MEC server with limited resources. An efficient and predictive computation offloading scheme is designed based on GT, where each vehicle fulfills its computation task either locally or remotely. The utility of vehicles is modeled by the mobility of vehicles, the processing time of the computation tasks, and the offloading costs. By giving NE, vehicles make optimal offloading decisions to execute tasks locally or offload them to direct or $j$-hop-away MEC servers.

To address the load balancing of computation resources at the edge servers, Zhang et al. \cite{zhang2019task} proposed a load balancing task offloading scheme based on two games. Firstly,  a game-based offloading algorithm is designed to optimize vehicles' task offloading decisions in each time slot. Secondly, the authors proposed a predictive GT-based task offloading algorithm to estimate the vehicles' locations by estimating their tasks processing time, so that the tasks can be executed at the MEC servers where the vehicles will arrive. 

Gu and Zhou \cite{gu2019task} focused on the task offloading in vehicular MEC environment with incomplete information (e.g., when channel states and vehicle mobility are not completely available).  Firstly, both MEC servers and vehicles with excess computation capacities are viewed as edge nodes. Besides, the interactions between tasks and edge nodes are modeled as a matching game that aims to minimize the average delay of task offloading and satisfy the energy consumption of edge nodes.
	
	\begin{itemize}
		\item  \textbf{\textbf{Joint task offloading and resource allocation game}}
	\end{itemize}
	
	Zhao et al.  \cite{zhao2019computation} proposed a collaborative task offloading and resource allocation scheme based on MEC and traditional Cloud Computing (CC). The network consists of a CC server, an MEC server with computing resource $ F $, $ N $ vehicles,and $ M $ RSUs located along the highway with the same coverage $ R  $. The offloading decisions are made by a game-theoretic approach. In the game, each vehicle makes decisions to process the computation tasks locally, offload them to the MEC server through an RSU, or offload them to the cloud server. The utility of each vehicle is modeled by metrics of task processing delay, cost for resource computing, and the normalized factors. The optimal computation offloading strategy (based on NE) and computation resource allocation are obtained by designing a distributed algorithm that works iteratively between decisions of computation offloading and resource allocation.
	
	Considering that there exists competition among MEC servers when deciding their optimal resource price to make more profit, an optimal multilevel offloading scheme \cite{zhang2017optimal} is designed by formulating the interaction between vehicles and MEC servers as a Stackelberg game. In this game, the MEC servers are the leaders who dynamically assigned computing resources to vehicles through an incentive method.  MEC servers' strategies are the amount of resources bought from the resource pool and the price of their resources sold to vehicle players. On the other hand, vehicles act as followers who make the optimal offloading decisions by dynamically reacting to the resource prices advertised by MEC servers. The existence of the NE is proved, and an optimal offloading algorithm is designed to maximize the revenue of the service providers and meet the delay constraints of the tasks.
	
	The security protection in the VEC  environment is addressed in \cite{huang2017distribute} by designing a reputation management scheme for the SP to allocate resources to vehicles based on their reputations. An SP scheduled multiple MEC servers to process computation offloading requests of vehicles; each MEC server runs a Virtual Machine (VM). For allocating computation resources, a bargaining game is formulated where the SP acts as a decision-maker who sets resource budgets based on the reputation and requirements of vehicles. Furthermore, each vehicle determines the amount of offloaded computations according to the budgets issued by the SP. This scheme not only guarantees optimal computation offloading for vehicles but also improves the detection of misbehaviors with low detection rates.

	An energy-efficient VEC framework is developed by integrating big data with VEC \cite{zhou2018begin}. A Stackelberg game-based energy-efficient resource allocation strategy is designed to solve the workload offloading problem. Similar to the scheme in \cite{huang2017distribute}, the game is played by edge computing SPs and vehicles in two stages. In the first stage, the RSU issues its computing service price based on the service revenue. In the second stage, each vehicle optimizes its offloading proportion based on the issued price and the edge computing time and electricity cost. This scheme guarantees energy efficiency and satisfies both QoS and Quality of Experience (QoE). Moreover, handover for MEC switching connections among RSUs is also considered to enhance service delivery capability.
      
\begin{itemize}
	\item  \textbf{\textbf{Edge content dissemination game}}
\end{itemize}

      In \cite{hui2019edge}, the authors developed an auction game-based two-stage model for edge content dissemination in urban VNs. Firstly, the provider uploads the content to the Edge Computing Devices (ECD), where the content is cached temporarily. Then a two-stage game is designed for ECD to select the optimal relay. In the first stage, a first-price sealed-bid auction game is developed for ECD to select relay vehicles with high transmission capability. Then the cached contents are delivered to these candidate vehicles through V2I communication. In the second stage, these candidate relays transmit the contents to the destination vehicle. Simulation results show that this incentive content dissemination scheme outperforms conventional methods.
      
		\begin{table*}  
		\caption{Summary of games for MEC-implemented VEC}
		\label{tab_MEC}
        \scriptsize
		\renewcommand*{\arraystretch}{.6}
		\begin{center}
			\begin{tabular}{|m{.018\textwidth}|m{.16\textwidth}|m{.055\textwidth}|m{.06\textwidth}|m{.13\textwidth}|m{.12\textwidth}|m{.07\textwidth}|m{.001\textwidth}|m{.001\textwidth}|m{.001\textwidth}|m{.001\textwidth}|m{.001\textwidth}|m{.001\textwidth}|m{.001\textwidth}|m{.001\textwidth}|}
				\hline
					\multirow{7}{2cm}{\textbf{Ref}}&\multirow{7}{1cm}{\textbf{Objective}}&\multicolumn{4}{c|}{\multirow{4}{1cm}{\textbf{Game}}}&\multirow{7}{1cm}{\textbf{Servers}}
					 &\multirow{7}{2pt}{\rotatebox[origin=c]{90}{\textbf{V2V}}}&\multirow{7}{2pt}{\textbf{\rotatebox[origin=c]{90}{\textbf{V2I}}}}&\multirow{7}{1cm}{\rotatebox[origin=c]{90}{\textbf{Mobility}}}&\multirow{7}{1cm}{\rotatebox[origin=c]{90}{\textbf{QoS}}}&\multirow{7}{1cm}{\rotatebox[origin=c]{90}{\textbf{Security}}}&\multirow{7}{1cm}{\rotatebox[origin=c]{90}{\textbf{Energy}}}&\multirow{7}{1cm}{\rotatebox[origin=c]{90}{\textbf{Cost}}} &\multirow{7}{1cm}{\rotatebox[origin=c]{90}{\textbf{Continuity}}} \\
					&&\multicolumn{4}{c|}{}&&&&&&&&&\\	 
					&&\multicolumn{4}{c|}{}&&&&&&&&&\\	 
					&&\multicolumn{4}{c|}{}&&&&&&&&&\\	 
			     \cline{3-6}
					&&\multirow{3}{1cm}{Type}&\multirow{3}{1cm}{Player}&\multirow{3}{1cm}{Strategy}&\multirow{3}{1cm}{Utility}&&&&&&&&&\\
					&&&&&&&&&&&&&&\\	 
					&&&&&&&&&&&&&&\\
				\hline
				 \multicolumn{15}{|c|}{\textbf{Computation or Task Offloading  Game}}\\
				 \hline
					\cite{liu2018computaion} &\begin{itemize}[leftmargin=3pt]\setlength{\itemsep}{0pt}\item Decrease computation overhead \item Select a proper wireless channel for each vehicle\end{itemize}&Potential Game&Vehicles&Wireless channel selection for offloading&Uplink data rate&MEC servers &&$\surd$&&$\surd$&&&&\\
              \cline{2-15}    
	                \cite{zhang2017mobile}&\begin{itemize}[leftmargin=3pt]\setlength{\itemsep}{0pt}\item Reduce latency and offloading cost \item Make optimal offloading decisions \end{itemize} & Non-cooperative game&Vehicles&Execute tasks locally or offload them to MEC servers& \begin{itemize}[leftmargin=5pt]\setlength{\itemsep}{0pt}\item Average offloading costs of vehicles \item Tasks processing delay \end{itemize}\item&\begin{itemize}[leftmargin=3pt]\setlength{\itemsep}{0pt}\item RSUs with MEC servers \item A cloud server\end{itemize}&$\surd$&$\surd$&$\surd$&$\surd$&&$\surd$&$\surd$&\\
	           \cline{2-15}
                   	 \cite{zhang2019task}&\begin{itemize} [leftmargin=3pt]\setlength{\itemsep}{0pt} \item Load balancing \item Minimize the processing delay of all the computation tasks under delay constraints \item Jointly optimize the selection of MEC servers and vehicles’ task offloading decisions \end{itemize}&Non-cooperative game&Vehicles&\begin{itemize} [leftmargin=5pt] \setlength{\itemsep}{0pt}\item Offloading decision \begin{itemize}[leftmargin=3pt]\setlength{\itemsep}{0pt}\item Computes locally\item Offloads to MEC server \item Offloads to cloud server\end{itemize}\end{itemize} &Task’s execution time and data transmitting time&\begin{itemize}[leftmargin=3pt]\setlength{\itemsep}{0pt}\item RSUs with MEC servers \item A cloud server\end{itemize}&$\surd$&$\surd$&$\surd$&$\surd$&&&&\\     
     		    \cline{2-15}
             		  \cite{gu2019task} &\begin{itemize} [leftmargin=5pt] \setlength{\itemsep}{0pt}\item Minimize the task offloading delay  \item Satisfy energy consumption of vehicles and edge nodes \end{itemize}&Matching game&Task and edge nodes&Task matching &\begin{itemize}[leftmargin=3pt]\setlength{\itemsep}{0pt}\item Task: offloading delay\item Edge nodes: energy consumption \end{itemize}&\begin{itemize}[leftmargin=5pt]\item RSUs with MEC servers\item Vehicular edge nodes\end {itemize}&$\surd$&$\surd$&$\surd$&$\surd$&&$\surd$&$\surd$&\\
                 \hline  	
                 	 \multicolumn{15}{|c|}{\textbf{Joint Task Offloading and Resource Allocation Game}}\\
                 \hline 
                        \cite{huang2017distribute} &\begin{itemize} [leftmargin=3pt] \setlength{\itemsep}{0pt}\item Provide security protection using reputation management \item Optimize resource allocation for service providers \item Optimize computation offloading for vehicles  \end{itemize}&Bargaining game&SP with MEC servers and vehicles &\begin{itemize}[leftmargin=3pt]\setlength{\itemsep}{0pt}\item SP: Resource budgets \item Vehicles: Amount of requesting computation resources \end{itemize}&\begin{itemize}[leftmargin=3pt]\setlength{\itemsep}{0pt}\item SP: Sum of resource budgets \item Vehicles: total processing delay of computation and cost for service\end{itemize}&\begin{itemize}[leftmargin=3pt]\setlength{\itemsep}{0pt}\item RSUs with MEC servers \item BSs with MEC servers \end{itemize}&$\surd$&$\surd$&&$\surd$&$\surd$&&$\surd$&\\
                \cline{2-15}
	          	 	 \cite{zhao2019computation}& \begin{itemize}[leftmargin=3pt]\setlength{\itemsep}{0pt}\item Optimize computation offloading for vehicles \item Optimize resource allocation of MECs and cloud\item Decrease the system complexity without loss of the performance\end{itemize}&Non-cooperative game &Vehicles&\begin{itemize}[leftmargin=3pt]\setlength{\itemsep}{0pt}\item Computes locally\item Offloads to MEC server \item Offloads to cloud server\end{itemize}&\begin{itemize}[leftmargin=3pt]\setlength{\itemsep}{0pt}\item Task processing delay \item Cost of computation resource \item Normalization factor \end{itemize}&\begin{itemize}[leftmargin=3pt]\setlength{\itemsep}{0pt}\item A MEC server \item A cloud server\end{itemize}&&$\surd$&$\surd$&$\surd$&&&&\\
	            \cline{2-15}
			         \cite{zhang2017optimal} &  \begin{itemize}[leftmargin=3pt]\setlength{\itemsep}{0pt}\item Optimal resource allocation \begin{itemize} [leftmargin=10pt]\setlength{\itemsep}{0pt}\item Backup resources sharing among MEC servers \item MEC servers dynamically  assign their resources to vehicles \end{itemize} \item Satisfy the delay constraints of tasks \end{itemize}&Stackelberg game&MEC servers and vehicles& \begin{itemize}[leftmargin=3pt] \item\setlength{\itemsep}{0pt} Vehicles: Choose the offloading target MEC servers \item MEC servers: \begin{itemize} [leftmargin=3pt] \item the amount of the resources bought from the BCS \item price of their resources sold to vehicles \end{itemize} \end{itemize}&\begin{itemize}[leftmargin=3pt] \setlength{\itemsep}{0pt}\item Vehicles: total time cost task offloading \item MEC servers: revenue of the resources\end{itemize}& \begin{itemize}[leftmargin=3pt]\setlength{\itemsep}{0pt}\item RSUs with MEC server \item A cloud server\end{itemize}&&$\surd$& &$\surd$&&&$\surd$&\\
			 \cline{2-15}
					 \cite{zhou2018begin}&\begin{itemize}[leftmargin=3pt]\setlength{\itemsep}{0pt}\item Dynamic resources allocation to guarantee: \begin{itemize}[leftmargin=8pt]\setlength{\itemsep}{0pt}\item Energy efficiency \item QoS guarantee \item QoE guarantee \end{itemize} \item Optimal computation offloading \end{itemize}&Stackelberg game&SPs and vehicular users&\begin{itemize}[leftmargin=3pt]\setlength{\itemsep}{0pt}\item SP: Service provision revenue and electricity consumption \item Vehicles: proportion  of resources to offload\end{itemize}&\begin{itemize}[leftmargin=3pt]\setlength{\itemsep}{0pt}\item SP: service provision revenue \item Vehicles: edge computing time and electricity cost \end{itemize}&\begin{itemize}[leftmargin=3pt]\setlength{\itemsep}{0pt}\item RSUs with edge controllers \item A cloud server \end{itemize}&$\surd$&$\surd$&$\surd$&$\surd$&&$\surd$&$\surd$&$\surd$\\
			 \hline
			 \multicolumn{15}{|c|}{\textbf{Edge Content Dissemination Game}}\\
		 	\hline
					\cite{hui2019edge}&Select the relay vehicles to satisfy different transmission requirements&Auction game& ECD and vehicles&\begin{itemize} [leftmargin=3pt]\setlength{\itemsep}{0pt}\item ECD: Candidate relay vehicles selection \item Relay vehicles: Optimal bid selection \end{itemize}&\begin{itemize} [leftmargin=3pt]\setlength{\itemsep}{0pt}\item ECD: Transmission capabilities of the\item  Vehicles: Bid and relay cost
				\end{itemize}&RSUs with ECDs&$\surd$&$\surd$&$\surd$&$\surd$&&&$\surd$&\\
		  \hline		
			\end{tabular}
		\end{center}
	\end{table*}

\subsubsection{FC-implemented VEC}
Like MEC, FC can also enable edge computation by implementing a decentralized computing infrastructure based on heterogeneous and collaborative fog nodes (e.g., access points, routers, and IoT gateways) placed at any point in the architecture between the end-user devices and the near-user edge devices  \cite{dolui2017comparison}.  As vehicular fog is also a resource-intensive component with limited resources, GT can be used for proper and fair computation and resource allocation for Vehicular Fog Computing (VFC) networks in recent researches. Table. \ref{tab_FC} summarizes the  GT-based schemes of computation or task offloading and resources allocation in MEC-implemented VEC

Zhang et al. \cite{zhang2017resource} modeled the resource allocation in the software-defined VFC network as a mean-field game. The Fog-Small BSs (F-SBSs) are seen as players aiming to optimize the power control according to the payoff functions. The payoff function of an F-SBS is formulated by metrics of downlink capacity, energy, interference, and caching reward. The game can help vehicles to choose the satisfying  F-SBSs with optimal power control. Furthermore, Sutagundar et al. \cite{sutagundar2019resource} proposed a game-based resource allocation algorithm for fog enhanced vehicular services. Each fog has multiple VMs which act as players in the game and compete for resources to provide service for the requested vehicles. The algorithm enables proper resource allocation for vehicles within one hop of the RSU; once the vehicle gets connected to the next RSU, the VM migrates from one fog instance to another to provide the continuity of service.

In \cite{xu2018low}, a VFC framework with low-latency and massive-connectivity is designed to relieve the insufficient computation capability of BS and reduce the processing delay. A two-sided matching game is designed for the task assignment, where vehicles with idle resources act as Vehicular Fog Nodes (VFNs). VENs can share their resources so that computation tasks of User Equipment (UEs) can be offloaded to these fog nodes. By combining contract and matching theories, Zhou et al.  \cite{zhou2019computation} further developed a two-stage computation resource allocation and task assignment scheme to minimize the delay and guarantee QoS and QoE for UEs. In the first stage, an incentive model is designed motivate vehicles to sign the contract of resource sharing. In the second stage, vehicles negotiate an agreement with the BSs in the previous step serve as VFNs. Then a pricing-based stable matching is designed for resources assignment between UEs and VFNs based on the dynamic preference lists. 

In \cite{klaimi2018theoretical}, parked vehicles perform as fogs to allocate resources for applications of local mobiles with different QoS requirements. Firstly, the application demands are classified into high priority (HP) demands and low priority (LP) demands. Then the parked vehicles play a distributed potential game to efficiently allocate the fog resources to satisfy the applications' demands while minimizing the latency and the number of redirected requests to the cloud.

Location privacy-preserving in VFC networks is studied in \cite{kang2017privacy}. The authors presented \textit{ pseudonym fogs}, each of which consists of a set of roadside infrastructures and is deployed nearby vehicles.  These pseudonym fogs play a real-time pseudonym change game interactively to allocate pseudonym resources to nearby vehicles. This scheme improves the vehicles' location privacy with reduced pseudonym management overheads. One disadvantage of this scheme is that it does not apply to situations with sparse vehicles.

\begin{table*} 
	\scriptsize   
	\caption{Summary of games for FC-implemented VEC}
	\label{tab_FC}
	\renewcommand*{\arraystretch}{.6}
	\begin{center}
		\begin{tabular}{|m{.018\textwidth}|m{.14\textwidth}|m{.05\textwidth}|m{.05\textwidth}|m{.14\textwidth}|m{.1\textwidth}|m{.09\textwidth}|m{.001\textwidth}|m{	.001\textwidth}|m{.001\textwidth}|m{.001\textwidth}|m{.001\textwidth}|m{.001\textwidth}|m{.001\textwidth}|m{.005\textwidth}|m{.001\textwidth}|}
			\hline
				\multirow{7}{2cm}{\textbf{Ref}}&\multirow{7}{1cm}{\textbf{Objective}}&\multicolumn{4}{c|}{\multirow{4}{1cm}{\textbf{Game}}}&\multirow{7}{1cm}{\textbf{Servers}}
			&\multirow{7}{2pt}{\rotatebox[origin=c]{90}{\textbf{V2V}}}&\multirow{7}{2pt}{\textbf{\rotatebox[origin=c]{90}{\textbf{V2I}}}}&\multirow{7}{2pt}{\textbf{\rotatebox[origin=c]{90}{\textbf{V2P}}}}&\multirow{7}{1cm}{\rotatebox[origin=c]{90}{\textbf{Mobility}}}&\multirow{7}{1cm}{\rotatebox[origin=c]{90}{\textbf{QoS}}}&\multirow{7}{1cm}{\rotatebox[origin=c]{90}{\textbf{Security}}}&\multirow{7}{1cm}{\rotatebox[origin=c]{90}{\textbf{Energy}}}&\multirow{7}{1cm}{\rotatebox[origin=c]{90}{\textbf{Cost}}} &\multirow{7}{1cm}{\rotatebox[origin=c]{90}{\textbf{Continuity}}} \\
			&&\multicolumn{4}{c|}{}&&&&&&&&&&\\	 
			&&\multicolumn{4}{c|}{}&&&&&&&&&&\\	 
			&&\multicolumn{4}{c|}{}&&&&&&&&&&\\	 
			\cline{3-6}
			&&\multirow{3}{1cm}{Type}&\multirow{3}{1cm}{Player}&\multirow{3}{1cm}{Strategy}&\multirow{3}{1cm}{Utility}&&&&&&&&&&\\
			&&&&&&&&&&&&&&&\\	 
			&&&&&&&&&&&&&&&\\		
			\hline
					 \cite{zhang2017resource}&\begin{itemize} [leftmargin=3pt]\setlength{\itemsep}{0pt}\item Optimal serving point selection \item Optimal power control of small base stations\end{itemize}&Mean-field game&Fog small base stations&Transmit power& \begin{itemize} [leftmargin=3pt]\setlength{\itemsep}{0pt}\item Value function  \item energy \item interference \item downlink capacity \item catching reward \end{itemize}&	\begin{itemize} [leftmargin=3pt]\setlength{\itemsep}{0pt}\item A macro BS with fog server \item Small BSs with fog servers \item a cloud server\end{itemize}&&$\surd$&&&$\surd$&&$\surd$&$\surd$&\\
			 \hline
			 		\cite{sutagundar2019resource}&\begin{itemize} [leftmargin=3pt]\setlength{\itemsep}{0pt}\item Prediction of resources required and the
					availability of resources \item Fair and efficient fog resources allocation for vehicles  \end{itemize}&Non-cooperative game&VMs&The amount of resources allocated&Estimation of amount of required resources&\begin{itemize} [leftmargin=3pt]\setlength{\itemsep}{0pt}\item RSUs with VM-based fog servers \item A cloud server\end{itemize}&&$\surd$&&$\surd$&$\surd$&&&$\surd$&$\surd$\\
			\hline
					 \cite{xu2018low}&\begin{itemize} [leftmargin=3pt]\setlength{\itemsep}{0pt}\item Relive overload on BSs \item Reduce total network delay \end{itemize}&Matching game&UEs and vehicles&UE-VFN pair matching&\begin{itemize} [leftmargin=3pt]\setlength{\itemsep}{0pt}\item Total delay \item Price for using the resources of VFN \end{itemize}& Vehicular fog nodes& &&$\surd$&$\surd$&$\surd$&&&&\\		
			\hline
					\cite{zhou2019computation}&\begin{itemize} [leftmargin=3pt]\setlength{\itemsep}{0pt}\item Minimize total delay \item Motivate vehicles to share resources using contract incentive \item Optimal task assignment \end{itemize}&Matching game&UEs and vehicles&Pricing-based UE-VFN pair matching&Expected social welfare&Vehicular fog nodes&&&$\surd$&$\surd$&$\surd$&&&$\surd$&\\
		   \hline
		   			 \cite{klaimi2018theoretical} &\begin{itemize} [leftmargin=3pt]\setlength{\itemsep}{0pt}\item Minimize the latency and optimize the  utilization of computation and energy \item Allocate fog resources dynamically  \end{itemize}&Potential game&Fog vehicles&\begin{itemize} [leftmargin=3pt]\setlength{\itemsep}{0pt}\item Do not satisfy any type of demands \item Satisfy only HP demands \item Share the computation \item Satisfy only LP demands  \end{itemize}&\begin{itemize} [leftmargin=3pt]\setlength{\itemsep}{0pt}\item CPU and energy utilization \item Task processing time \end{itemize}&Parked fog vehicles&&&$\surd$&&$\surd$&&$\surd$&$\surd$&\\
		   	\hline
		   			\cite{kang2017privacy} &Secure pseudonym changing management with low overhead&Non-cooperative game&Vehicles&Change pseudonym or maintain current pseudonym&\begin{itemize} [leftmargin=3pt]\setlength{\itemsep}{0pt}\item Vehicle-side entropy \item Pseudonym change cost\end{itemize}& \begin{itemize} [leftmargin=3pt]\setlength{\itemsep}{0pt}\item Pseudonym fogs with roadside infrastructures\item A cloud server\end{itemize}&$\surd$&$\surd$&&$\surd$&$\surd$&$\surd$&&$\surd$&\\
             \hline
		\end{tabular}
	\end{center}
\end{table*}

\subsubsection{ Cloudlet-implemented VEC} 
As is shown in Fig. \ref{fig_vec}, a cloudlet in VNs is a trusted cluster of roadside infrastructures (e.g., RSU and BS) with resources available to use for vehicles near form them \cite{satyanarayanan2009case}. The infrastructure nodes in a cloudlet can share resources; different cloudlets can coordinate and cooperate for communication. Table. \ref{tab_Cloudlet} summarizes the recent researches employing GT in cloudlet-implemented VNs.

To operate the resource allocation of cloudlets, Yu et al. \cite{yu2016optimal} defined a new paradigm of 5G-enabled VNs, an enhanced cloud radio access network, which integrates geo-distributed cloudlets with SDN and D2D technologies. The authors exploited the matrix GT for cloudlet resource allocation and gave explicit solutions to global optimization. The study in  \cite{tao2017resource} designs an RSU cloud resource allocation scheme based on a non-cooperative game. To deal with the low efficiency of the convergence at NE and to achieve the near Pareto-optimal flow allocation, the previous one-shot game is extended to a repeated game where the punishment of misbehaviors is considered to avoid selfish vehicles deviating from the cooperation.

The authors of \cite{yu2015cooperative} proposed a coalition game model to stimulate cloud SPs cooperatively form coalitions to share their idle resources. Firstly, each SP evaluates its revenue and decides whether to join a coalition in the cloud. Then in each coalition, the two types of players act as 1) the $inviters$ who would like to rent resources $R_{from}$ and 2) the $invitees$ who would like to lease out their resources $R_{to}$. By using a two-sided matching game, the $inviters$ and $invitees$ match their demands effectively and form stable coalitions cooperatively. The coalition is proved to be $\mathbb{D}_c$ stable, and Pareto optimal solution is the only stable solution. 

Different from the works mentioned above, vehicles are viewed as a dynamic extension of cloudlets and share computation resources to cloudlets in \cite{lin2019vehicle}. The authors designed a Stackelberg game to stimulate vehicles to trade in their computation resources with a cloudlet SP. After obtaining the vehicles' states from the cloud server,  the cloudlet SP first sets the optimal discriminatory price for the trade based on the computation demands of the vehicles. Then each vehicle decides the computation amount it will buy from cloudlet SP according to the giving price. The existence and uniqueness of NE are proved, and the solution has good scalability for more participating vehicles.

\begin{table*} 
	\scriptsize    
	\caption{Summary of Games for Cloudlet-implemented VEC}
	\label{tab_Cloudlet}
	\renewcommand*{\arraystretch}{.4}
	\begin{center}
		\begin{tabular}{|m{.012\textwidth}|m{.15\textwidth}|m{.05\textwidth}|m{.08\textwidth}|m{.12\textwidth}|m{.15\textwidth}|m{.04\textwidth}|m{.001\textwidth}|m{	.001\textwidth}|m{.001\textwidth}|m{.001\textwidth}|m{.001\textwidth}|m{.001\textwidth}|m{.001\textwidth}|m{.001\textwidth}|}
			\hline
				\multirow{10}{2cm}{\textbf{Ref}}&\multirow{10}{1cm}{\textbf{Objective}}&\multicolumn{4}{c|}{\multirow{4}{1cm}{\textbf{Game}}}&\multirow{10}{1cm}{\textbf{Servers}}
			&\multirow{10}{2pt}{\rotatebox[origin=c]{90}{\textbf{V2V}}}&\multirow{10}{2pt}{\textbf{\rotatebox[origin=c]{90}{\textbf{V2I}}}}&\multirow{10}{1cm}{\rotatebox[origin=c]{90}{\textbf{Mobility}}}&\multirow{10}{1cm}{\rotatebox[origin=c]{90}{\textbf{QoS}}}&\multirow{10}{1cm}{\rotatebox[origin=c]{90}{\textbf{Security}}}&\multirow{10}{1cm}{\rotatebox[origin=c]{90}{\textbf{Energy}}}&\multirow{10}{1cm}{\rotatebox[origin=c]{90}{\textbf{Cost}}} &\multirow{10}{1cm}{\rotatebox[origin=c]{90}{\textbf{Continuity}}} \\
			&&\multicolumn{4}{c|}{}&&&&&&&&&\\	 
			&&\multicolumn{4}{c|}{}&&&&&&&&&\\	 
			&&\multicolumn{4}{c|}{}&&&&&&&&&\\	 
			\cline{3-6}
			&&\multirow{6}{1cm}{Type}&\multirow{6}{1cm}{Player}&\multirow{6}{1cm}{Strategy}&\multirow{6}{1cm}{Utility}&&&&&&&&&\\
			&&&&&&&&&&&&&&\\	 
			&&&&&&&&&&&&&&\\
				&&&&&&&&&&&&&&\\
					&&&&&&&&&&&&&&\\
						&&&&&&&&&&&&&&\\
			\hline
				\cite{yu2016optimal}&\begin{itemize} [leftmargin=3pt]\setlength{\itemsep}{0pt}\item Resource sharing  among the geo-distributed cloudlet \item Improve resource utilization and reduce power consumption \end{itemize}&Non-cooperative game&RSU cloudlets &The amount of resources allocated&Resources  Utilization \begin{itemize} [leftmargin=3pt]\setlength{\itemsep}{0pt}\item CPU resource \item memory resource \item bandwidth resource \end{itemize}& RSU cloudlets &&$\surd$&&$\surd$&&$\surd$&$\surd$&\\
			\hline
				 \cite{tao2017resource} &Effective flow rate assignment&Non-cooperative game&Vehicles&Flow rate allocated to each vehicle&Transmission efficiency&RSU cloudlets &&$\surd$&&$\surd$&&&$\surd$&\\
		   \hline
		       \cite{yu2015cooperative}&Stimulate cloud SPs to form coalitions for resources sharing& Coalition game & $inviters$ and $invitees$ &\begin{itemize} [leftmargin=3pt]\setlength{\itemsep}{0pt}\item $inviters$: rent resources \item $invitees$: lease out resources \end{itemize}& \begin{itemize} [leftmargin=3pt]\setlength{\itemsep}{0pt}\item $inviters$: \begin{itemize} [leftmargin=3pt]\setlength{\itemsep}{0pt}\item satisfaction function \item rent payment\end{itemize} \item $invitees$: \begin{itemize} [leftmargin=3pt,itemsep=0pt]\setlength{\itemsep}{0pt}\item satisfaction function\item rental income\end{itemize} \end{itemize}&SP cloudlets&$\surd$&$\surd$&&$\surd$&&&$\surd$&\\
		     \hline
			     \cite{lin2019vehicle}&Stimulate vehicles to share computation resources to cloudlets&Stackelberg game&A cloudlet SP and individual vehicles&\begin{itemize} [leftmargin=3pt]\setlength{\itemsep}{0pt}\item Cloudlet SP: discriminatory price\item Vehicles: computation trading amount\end{itemize}&\begin{itemize} [leftmargin=3pt]\setlength{\itemsep}{0pt}\item Cloudlet SP: cost of renting \item Vehicles: benefit of the trading\end{itemize}&SP coudlets&$\surd$&&$\surd$&$\surd$&&&$\surd$&$\surd$\\
			  \hline
	\end{tabular}
\end{center}
\end{table*}

\subsubsection{VC in VEC}
A variety of services are provided to vehicles in VNs, which may cause inefficient utilization of resources. For example, the resources of parked vehicles are idle, while the vehicles in congested areas compete for insufficient resources.  In this context,  VC has emerged as a promising solution to the exploitation of underutilized resources by integrating VNs with CC. However, resource management of VC is more challenging compared with traditional CC because high mobility of vehicles may lead to "cloud resource mobility". Moreover, communication in VC is more vulnerable to attacks than that in traditional VNs because a large number of users share the resources.  GT has been employed by recent studies in the decision-making of resource management or security protection for VC-implemented VEC. Table. \ref{tab_VC} summarizes the works applying GT in VC-implemented VEC.

The study in \cite{mekki2017proactive} focuses on a hybrid wireless network access scheme in VC networks based on an evolutionary game. Vehicles cooperate to form a temporary cloud to share the resources (i.e., being a consumer or resource provider) or act alone to access to the conventional cloud. Each vehicle can change its strategies from the conventional cloud to the VC or inversely according to its payoff. To model the evolution of vehicles' strategies, the authors developed an evolutionary game that allows vehicles to determine to access conventional or vehicular clouds.

The study of \cite{liwang2019game}  developed an opportunistic V2V computation offloading scheme based on a two-player Stackelberg-game. In the game, each vehicular acts either as an SP with idle computation resources or a requester who has a computation-intensive task that can be carried out locally or offloaded to nearby providers.  The payoff functions involve the metrics of vehicular mobility, V2V communication duration, computational capabilities, channel conditions, and service costs under both circumstances of complete information and incomplete information. The method can determine the appropriate offloading rate, optimal SP selection, and the ideal pricing strategies of SPs.

To handle the service management in VCs, Aloqaily et al. \cite{aloqaily2017fairness} proposed a cooperative distributed game between vehicular drivers and SPs. The game aims at maximizing the SP's social utility while ensuring the vehicle driver's QoE, taking into account the efficiency of the service allocation and fairness among SPs. Considering the different preferences of SPs and consumers, Brik et al. \cite{brik2018gss} proposed a distributed game-based approach to model the interaction between consumer vehicles and provider vehicles for effective service allocation. In this game, SPs rent out their resources, and each consumer vehicle selects the optimal SP vehicle who can provide him services with satisfied QoS and cost.

\begin{table*} 
	\scriptsize    
	\caption{Summary of games for VC-implemented VEC}
	\label{tab_VC}
	\renewcommand*{\arraystretch}{.5}
	\begin{center}
		\begin{tabular}{|m{.018\textwidth}|m{.15\textwidth}|m{.063\textwidth}|m{.08\textwidth}|m{.13\textwidth}|m{.15\textwidth}|m{.04\textwidth}|m{.0001\textwidth}|m{.0001\textwidth}|m{.0001\textwidth}|m{.0001\textwidth}|m{.0001\textwidth}|m{.0001\textwidth}|m{.0001\textwidth}|m{.0001\textwidth}|}
			\hline
				\multirow{7}{2cm}{\textbf{Ref}}&\multirow{7}{1cm}{\textbf{Objective}}&\multicolumn{4}{c|}{\multirow{4}{1cm}{\textbf{Game}}}&\multirow{7}{1cm}{\textbf{Servers}}
			&\multirow{7}{2pt}{\rotatebox[origin=c]{90}{\textbf{V2V}}}&\multirow{7}{2pt}{\textbf{\rotatebox[origin=c]{90}{\textbf{V2I}}}}&\multirow{7}{1cm}{\rotatebox[origin=c]{90}{\textbf{Mobility}}}&\multirow{7}{1cm}{\rotatebox[origin=c]{90}{\textbf{QoS}}}&\multirow{7}{1cm}{\rotatebox[origin=c]{90}{\textbf{Security}}}&\multirow{7}{1cm}{\rotatebox[origin=c]{90}{\textbf{Energy}}}&\multirow{7}{1cm}{\rotatebox[origin=c]{90}{\textbf{Cost}}} &\multirow{7}{1cm}{\rotatebox[origin=c]{90}{\textbf{Continuity}}} \\
			&&\multicolumn{4}{c|}{}&&&&&&&&&\\	 
			&&\multicolumn{4}{c|}{}&&&&&&&&&\\	 
			&&\multicolumn{4}{c|}{}&&&&&&&&&\\	 
			\cline{3-6}
			&&\multirow{3}{1cm}{Type}&\multirow{3}{1cm}{Player}&\multirow{3}{1cm}{Strategy}&\multirow{3}{1cm}{Utility}&&&&&&&&&\\
			&&&&&&&&&&&&&&\\	 
			&&&&&&&&&&&&&&\\
			\hline
				 \cite{mekki2017proactive} &Optimal wireless network access &Evolutionary game&Conventional vehicle and VC member&\begin{itemize} [leftmargin=3pt]\setlength{\itemsep}{0pt}\item Access to conventional cloud \item Cooperate to be member in VC\end{itemize}&\begin{itemize} [leftmargin=3pt]\setlength{\itemsep}{0pt}\item Conventional Vehicle: \begin{itemize} [leftmargin=3pt]\setlength{\itemsep}{0pt}\item Amount of data received \item Payment for the service\end{itemize} \item VC member: \begin{itemize} [leftmargin=3pt]\setlength{\itemsep}{0pt}\item Data downloaded from other VC members \item Handover cost\end{itemize}\end{itemize}& \begin{itemize} [leftmargin=3pt]\setlength{\itemsep}{0pt}\item VC with eNB \item A cloud server \end{itemize}&$\surd$&$\surd$&&$\surd$&&&$\surd$&$\surd$\\
			\hline
				 \cite{liwang2019game} &\begin{itemize} [leftmargin=3pt]\setlength{\itemsep}{0pt}\item Determine appropriate offloading rate of requestors \item Select the appropriate computation SP\item Identify the ideal pricing strategy for SP\end{itemize} &Stackelberg game&A task vehicle and one of the server vehicles&\begin{itemize} [leftmargin=3pt]\setlength{\itemsep}{0pt}\item Task vehicle: the amount of offloaded data\item Service vehicle: service price\end{itemize}& \begin{itemize} [leftmargin=3pt]\setlength{\itemsep}{0pt}\item Task vehicle:  \begin{itemize} [leftmargin=3pt]\setlength{\itemsep}{0pt}\item task processing delay \item payment for service \end{itemize} \item Service vehicle: service price\end{itemize}&VC with RSU&$\surd$&&$\surd$&$\surd$&&&$\surd$&\\
			\hline
				\cite{aloqaily2017fairness} &\begin{itemize} [leftmargin=3pt]\setlength{\itemsep}{0pt}\item Fair resources allocation \item QoS and QoE guarantee \end{itemize}&Non-ooperative game&Vehicular driver and service provider&Cooperate or reject to cooperate&Social welfare function&VC&$\surd$&&&$\surd$&&&$\surd$&\\
			\hline
				\cite{brik2018gss}&Select the optimal provider vehicles to meet the QoS of consumer vehicles &Non-cooperative game&Consumer vehicle and provider vehicle&\begin{itemize} [leftmargin=3pt]\setlength{\itemsep}{0pt}\item Consumer vehicle: consume or not \item  Provider vehicle: offer or not  \end{itemize}&\begin{itemize} [leftmargin=3pt]\setlength{\itemsep}{0pt}\item Data throughput \item Successful execution ratio \item Execution duration \item Execution price\end{itemize}&VC&$\surd$&&$\surd$&$\surd$&&$\surd$&&\\
			\hline
		\end{tabular}
	\end{center}
\end{table*}

\subsection{Game Theory in Heterogeneous Vehicular Networks }
\begin{figure}[!hbt] 
	\centering
	\includegraphics[width =3.4in]{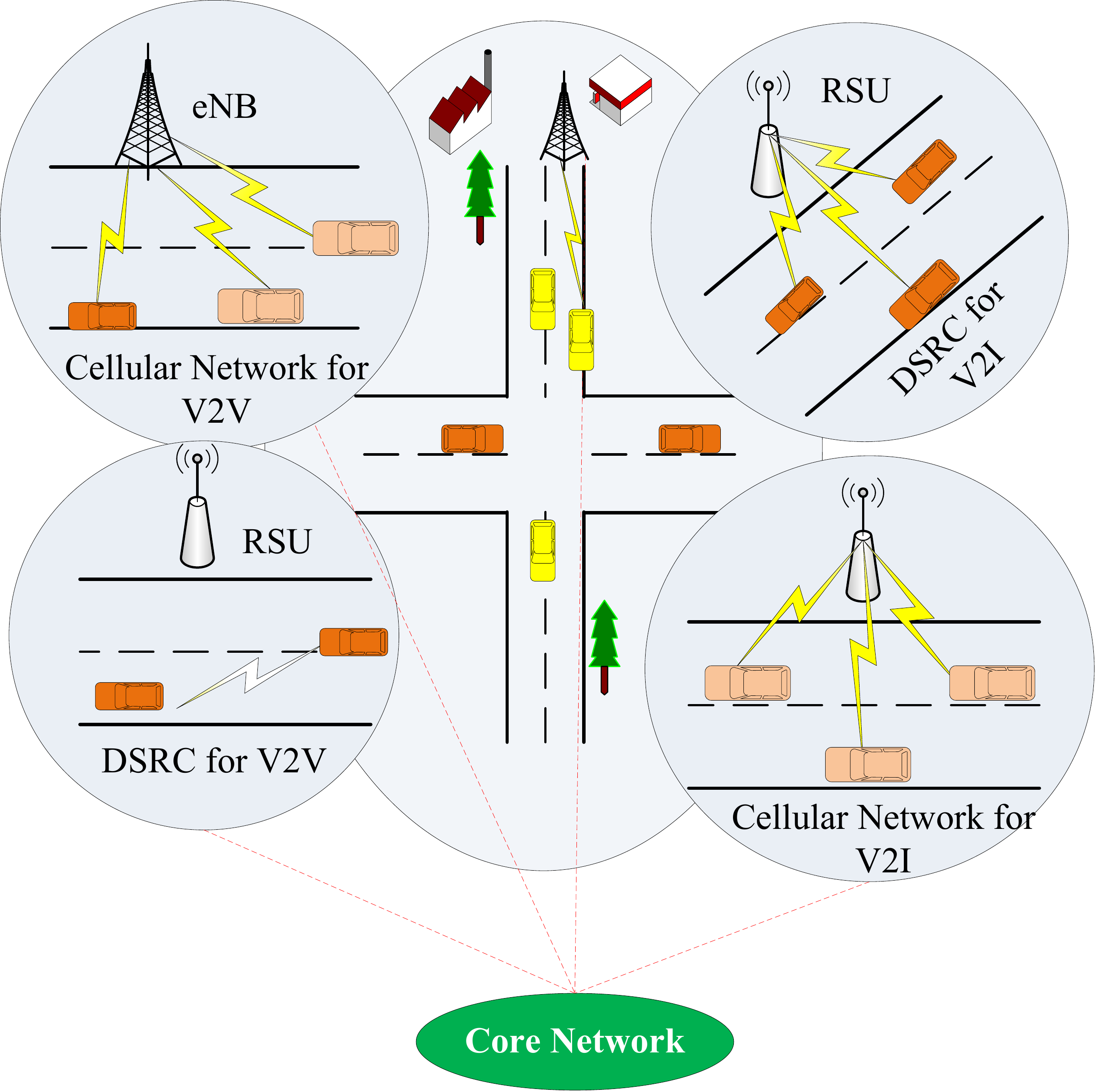}
	\caption{The architecture of HetVNETs \cite{zheng2015heterogeneous}}
	\label{fig_hetVN}
\end{figure}

The  HetVNETs apply the heterogeneous access technologies to provide ubiquitous connections to vehicles. As is shown in Fig.\ref{fig_hetVN},  HetVNETs provide both V2I and V2V communications with Long Term Evolution (LTE) and Dedicated Short-Range Communication (DSRC) technologies, respectively. The dynamic and instant integration of different technologies pose challenges to HetVNETs, such as resource management \cite{wang2014cellular,xiao2018spectrum}, efficient channel access \cite{mabrouk2016meeting,hui2017optimal,hui2019game,zhao2019optima}, and security protection \cite{sedjelmaci2017predict,chen2017congestion}.

\subsubsection{Resource allocation} Recently, CR technology has been integrated into HetVNETs for efficient utilization of spectrum resources \cite{wang2014cellular}. Xiao et al. \cite{xiao2018spectrum} investigated the radio resource allocation in HetVNETs by designing a game-theoretical approach. The downlink resource allocation for moving vehicles linking to different RSUs is modeled as a $ n $-person non-cooperative game. The game aims to optimize resource allocation and mitigate interference dynamically in terms of the interference among RSUs and the power constraints of RSUs. The fast-convergent equilibrium of the game is obtained, with which the performance of the HetVNETs can be improved.

\subsubsection{Optimal Access Control}  Optimizing the connection for vehicles with the lowest cost is challenging in HetVNETs, where different access technologies cause different costs in terms of download latency and bandwidth utilization. An always-best-connected paradigm \cite{mabrouk2016meeting} based on the signal game is designed to help vehicles in HetVNETs connect to the optimal network with a low connecting cost. However, this approach only considers a two-player competition case in a limited geographical area in urban scenarios. Hui et al. \cite{hui2017optimal,hui2019game} proposed an optimal access control for vehicles using a coalitional game. This method formulates the cooperation among vehicles based on their interests (contents cached in vehicles) and requests (contents to be downloaded). Vehicles in the same coalition can download the requested contents cooperatively with the minimum costs. 

Although the researches mentioned above ensure the connection of vehicles in HetVNETs, they do not consider the unexpected massive handoffs caused by constant changes of networks. However,  the facts such as crowded mobile vehicles, high sensitivity to delay, and packet loss make the traditional vertical handoff strategies ineffective in HetVNETs. To solve this problem,  \cite{zhao2019optimal} et al. proposed an optimal non-cooperative game approach for network selection in HetVNETs where performance parameters of networks are changing. Vehicles in the game attempt to switch to the network with higher evaluation, and the probabilistic strategies drive them toward a final stable convergence state.

\subsubsection{Security} 
Security in HetVNet is a challenging issue due to the vital exchanged information such as congestion, accidents, hazards, etc. Sedjelmaci .et al. \cite{sedjelmaci2017predict} designed a new attack detection and prediction scheme based on GT to detect and predict the misbehavior of a malicious vehicle in HetVNETs. The attack–defense problem is first formulated as a game between the attacker (i.e., misbehavior vehicle) and the services center. Furthermore, the future behavior of monitored vehicles can be predicted based on the NE. To alleviate the channel congestion on intersections in HetVNETs,  Chen et al. \cite{chen2017congestion} proposed a congestion game-theoretical transmission control scheme with a new concept RoS (Requirement of Safety) used to measure the security level of services.

\subsection{Game Theory in SDN-based VNs}
By decoupling the network management from message transmission, SDN technology facilitates the efficient utilization of network resources \cite{sun2015intelligent}. Existing SDN technologies include three types of structures: VN-based, cellular network-based, and hybrid architectures, among which hybrid structure can leverage the uncertain latency of ad-hoc networks and the cost of cellular networks. Li et al. \cite{li2016control} focused on leveraging the latency requirement of vehicles and the costs of cellular networks. A two-period Stackelberg game is designed for the interaction between the controller (leader) and vehicles (follower). The NE of the game gives the best strategies of the controller and vehicles, which are the optimal bandwidth allocation and the number of packets to be sent through the cellular network, respectively. Besides, an intelligent network selection scheme for data offloading is designed in \cite{aujla2017data} using a single-leader multi-follower Stackelberg game. The authors of \cite{chahal2019network} also employed a two-stage Stackelberg game for the optimal network selection in the SDN-based VNs with heterogeneous wireless interfaces. 

Considering the frequent handover when a lot of vehicles are connected to an RSU, Jia et al. \cite{jia2019bus} proposed a BUS-aided RSU connection scheme in SDN-IoV. The scheme employs an evolutionary game where the vehicles in the overlapped regions of the RSUs can adopt the optimal strategies to access to stable RSUs or BUSs. In  \cite{li2016network}, the assignment of network virtualization solutions in SDN-VNs is modeled as a non-cooperative game, of which the Pareto efficient solution is obtained. Alioua et al. \cite{alioua2019incentive} focused on the edge-caching problem in SDN-IoV by modeling the interaction between Content Provider (CP) and Multiple Network Operators (MNOs) as a Stackelberg game. The CP acting as a leader is responsible for publishing the quantity of content to be cached; the MNOs acting as followers respond to the leader subsequently based on caching price. 

\subsection{Game Theory in UAV-assisted VNs}

\begin{figure}[!hbt] 
	\centering
	\includegraphics[width =3.5in]{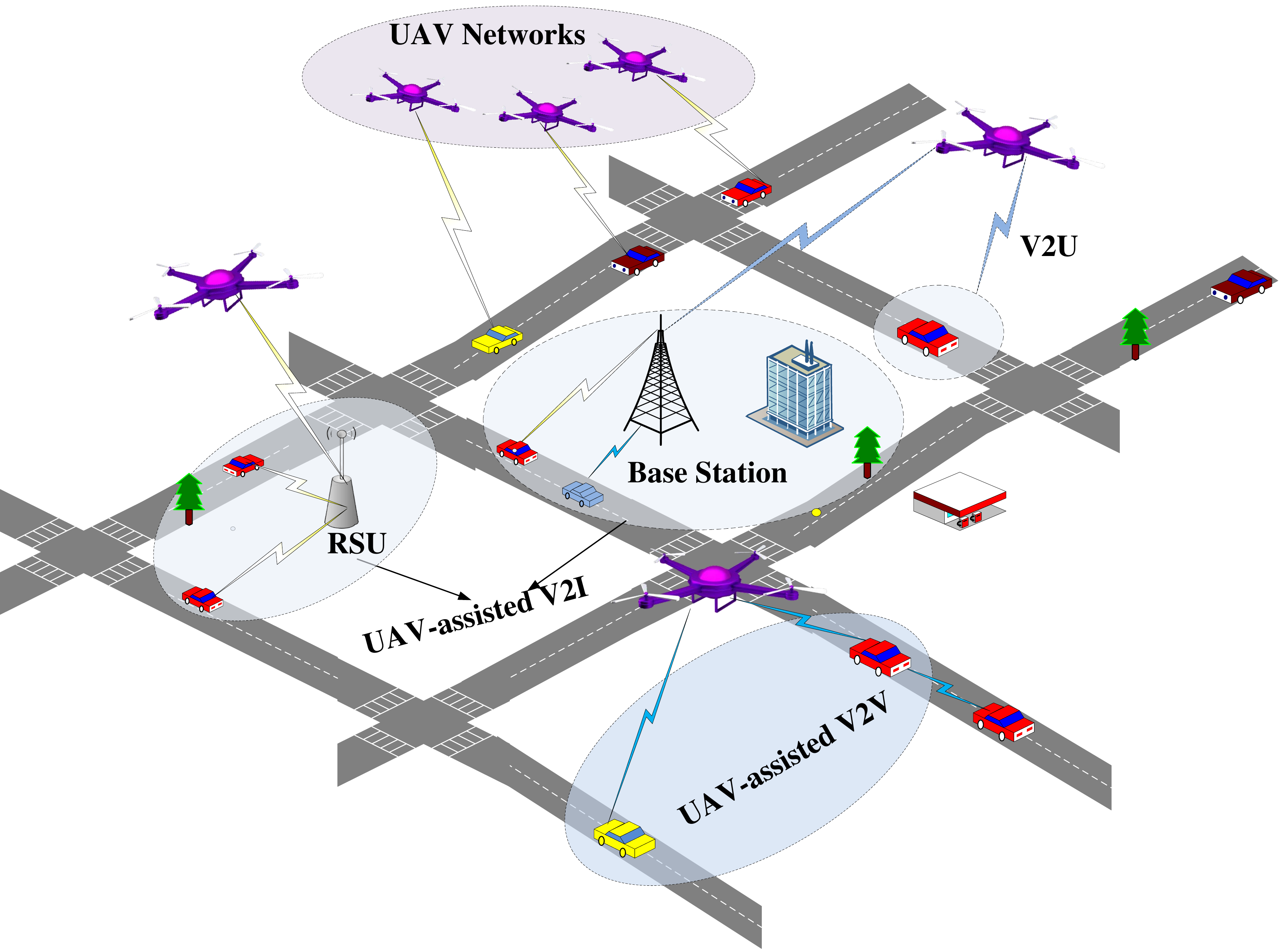}
	\caption{Architecture of UAV-assisted VNs \cite{shi2018drone}}
	\label{fig_uav}
\end{figure}

As an important and emerging element of IoT, UAV-assisted VNs can improve infrastructure coverage, enhance the network connectivity in resource-limited scenarios, and reduce the deployment cost. As is shown in Fig. \ref{fig_uav}, UAV enables three types of communication modes: UAV-assisted V2V communication, UAV-assisted V2I communication, and UAV networks \cite{shi2018drone}. However, due to the multihop communication from UAVs to infrastructures or vehicles and the dynamic of UAVs, the VNs may suffer QoS degradation and security threats. GT has been used in formulating decision-makings in security \cite{xiao2018uav,sedjelmaci2016intrusion} and QoS \cite{sedjelmaci2016intrusion,sedjelmaci2016intrusion} aspects in UAV-assisted VNs.

Xiao et al. \cite{xiao2018uav} proposed an anti-jamming relay game between the UAV and jammer for jamming resistance in the UAV-assisted VNs. In the game, the UAV determines whether to relay the OBU's messages to an RSU beyond the jammer, and the smart jammer attempts to attack by adjusting its jamming power. Moreover, NE is derived as a guide for a UAV to make the best UAV relay strategy for jamming resistance in VNs according to the channel condition and the transmission cost. The security problems of intrusion detection and attacker ejection in UAV-assisted VNs are discussed in \cite{sedjelmaci2016intrusion}. A Bayesian game is adopted to formulate the attack-defense interactions between the IDS and attackers to detect attacks with high accuracy and low overheads,

Besides security problems, GT is also applied in balancing QoS and resource consumption in UAV-assisted VNs. In \cite{alioua2018efficient}, the computation offloading is formulated as a two-player sequential game to balance the computation offloading delay and waste of energy in emergency situations in UAV-assisted VNs with less or no infrastructure. Besides, a mode selection scheme is proposed in \cite{wang2018mode} based on an evolutionary game. Each vehicle in the game selects the best communication mode from V2I, V2V, and Vehicle-to-UAV (V2U) in UAV-assisted VNs to maximize the transmission reliability while minimizing the cost of resource utilization.

\subsection{Game Theory in Other 5G-technology-based VNs}
Besides the above mentioned 5G-based technologies, GT is also employed in VNs integrating with other 5G-based emerging technologies. The content sharing problem in the D2D based V2X communication networks is studied in \cite{gu2016exploiting}, where the interaction between vehicles and eNBs is modeled as the stable matching game. Besides, to address the bandwidth slicing problem in SDN-based cellular networks, Zhou et al. \cite{zhou2018bandwidth} developed a hierarchical game framework to model the competitive interactions between internet service providers and users. Moreover, in \cite{perfecto2017millimeter}, GT is used in dynamically pair vehicles and optimize both transmission and reception beamwidths in Millimeter-Wave (mmWave) V2V communication. The congestion game is applied to solve the congestion in spectrum sharing VNs based on Licensed Shared Access (LSA) technology in  \cite{belghiti20185g}.

\section{Challenges}
	\label{sec_challenge}
	Possible future research directions of GT approaches related to VNs may consist of several aspects as follows:

\subsection{Handover Management in Game}

 GT has shown its potential for effective handover handling in VNs by modeling the network selection or resource management as games \cite{ kumar2016spectrum,jia2019bus,zhao2019optimal}. However, the massive handovers in next-generation VNs make the design of game models a challenging task. For example, vehicles may frequently switch among different types of networks (e.g., LTE and DSRC), move from one access point (e.g., RSU or edge) to another, or locate in the overlap among ranges of different networks or access points. Firstly, GT face the challenges of modeling 1) competition among vehicles for scarce resources, 2) cooperation among vehicles to form trustful coalitions (vehicles in the same coalition are connected to the same network or access point), and 3) coordination among access points and vehicles for effective resource allocation.  Furthermore, cooperative GT employed in seamless VM migration is an emerging solution for handovers but also a challenging task that requires effective interactions among access points.  On the one hand, sufficient prior knowledge about vehicles, especially the trajectory information, is a prerequisite for effective cooperation so that the content can be cached in advance along with the vehicle. On the other hand, the cooperation must be efficient with low complexity and communication overhead for real-time VM migration.

\subsection{Big Data Management in Game}

With the ever-increasing number of connections, information exchanges, and various applications, the data volume collected, transmitted, and processed in VNs has seen tremendous growth, resulting in the emergence of big data  \cite{cheng2018big}. The integration of temporal and spatial big data in the game-theoretical design of VNs remains an open issue.  Firstly, considering the resources of big data, the game will involve multiple types of players, such as vehicles, roadside infrastructures, and UAVs. What makes the problem more complicated is the recently proposed space-air-ground integrated VNs \cite{zhang2017software}, where the satellites are deployed in UAV-assisted VNs. Besides, the multi-tier (i.e., central clouds, edges, and vehicles) and heterogeneous networks (e.g., LTE and DSRC) incur large strategy spaces for the players. In this case, players should implement the information gathering and storing mechanism to make better decisions. Although storing more valuable data could increase the re-utilization of data, processing data with high-volume, high-velocity, and great-heterogeneity causes high storage costs and long delay Therefore, the trade-off between the cost of data processing and the value of big data should be considered in modeling the utility functions of the players. Mean-field game \cite{huang2010nce} could be a possible solution that provides a powerful mathematical tool for modeling the interaction among rational agents with massive exchanged information.

\subsection{Fair Resource Allocation}

Fairness is a critical metric of resource allocation (e.g., bandwidth, power, and frequency) in VNs to prevent resource starvation caused by unfair resource allocation. Although most studies focus on the effectiveness of resource allocation, the shared-medium, limited-resource, and high-mobility natures of VNs pose challenges to guarantee fairness among competitive users of resource allocation. Firstly, fair resource allocation in VNs potentially becomes more complex when the network suffers interference of multiple vehicles, especially in high-density scenarios. Secondly, with increasing demands for various applications with stringent QoS requirements in VNs, it is challenging to ensure both fairness and efficiency of resource allocation. Consequently, a trade-off between fairness and efficiency should be considered by designing fairness algorithms with acceptable computational complexity. Last, how to flexibly and fairly manage resource allocation in the dynamic and instant composition of different networks in HetVNETs is also a critical problem \cite{zheng2015heterogeneous}. Nash bargaining game has been used for fair resource allocation in wireless networks such as LTE networks \cite{huang2016pareto},  self-organizing heterogeneous networks \cite{senel2017fair}, and MEC \cite{zhu2018fair} networks. 



\subsection{Uncertain and Incomplete Information in VNs}
As is discussed before, most of the works modeling the rational behaviors of players (e.g.., vehicles and RSUs) in VNs as complete information games by assuming that the players know entirely other players' information. However, the knowledge acquired by players about their opponents is incomplete and uncertain in VNs, which leads to inaccuracy of game models. On the one hand, due to the highly dynamic networks, it is difficult for players to acquire complete information about their opponents. On the other hand, the security- or privacy-sensitive data is processed by mechanisms such as encryption and pseudonym, which can not be acquired by other players either. The application of GT with incomplete information in VNs is emerging to help players estimate the unknown information based on the belief models (i.e., probability distribution \cite{sun2019non}). Bayesian game is useful for scenarios with limited or highly variable context information and has been applied by a few studies. However, to build accurate beliefs on other players, the complex probabilistic model should be constructed, which causes large overheads and delay for the real-time vehicular communications. More studies are required to utilize GT with incomplete information to build cost-efficient and accurate beliefs on the participants in VNs.

\subsection{ Security in VNs}
As is discussed in this paper, the applications of GT for the security problems (e.g., intrusion detection, privacy-preserving, and trust management) in VNs mostly focus on modeling attackers, malicious or selfish nodes with assumed finite strategy spaces. However, it is not sufficient to protect the security of life-related communications in VNs due to the following limitations. Firstly, it is difficult for the defenders or legitimate players to obtain complete information about the payoffs and strategies of attackers. Although some studies apply the incomplete information game, it is an assumption of players' types or preferences. Secondly, one of the limitations of GT in modeling the players is that the agents are rarely entirely rational. Thirdly, the terms such as "security level", "privacy level", and "attacker strength" quantified manually with some equations may not be entirely accurate. Machine learning provides new insights for intelligent authentication in 5G \cite{fang2019machine} and malware detection schemes of IoT \cite{xiao2018iot}, which could be combined with GT for security strategy prediction and attacker model in the process of decision making in VNs. Especially, the Reinforce learning (RL) is a typical example that does not require players to know anything about other players \cite{tembine2018distributed}.

\subsection{ Efficiency of Game}

In non-cooperative games, NE is usually not efficient due to the lack of information about all players (i.e., each player only has the local information about a given player such as channel gains). Therefore, each participant gains an individual payoff selfishly. In contrast, cooperative games have information about all players in a coalition, where the joint payoff is shared among the members in the alliance. However, designing cooperative schemes is a challenging task. The players in a cooperate game should agree on an enforceable agreement before starting the game.  ”Threat” or ”punishment” is required for agreement mechanisms to force the players to comply with the commitment \cite{nash1953two}. Therefore, a central controller such as an RSU is required for collectin complete information and execute the enforced commitments and punishment policies. However, it could be challenging to satisfy this requirement in the environment where the deployment of roadside infrastructures is limited. Besides, cooperation stimulation and selfishness mitigation are critical but challenging when modeling the effective resource competition in VNs. Last but not least, the following aspects of GT should also be carefully considered in the game: 1) the result of GT should be precise and valid \cite{fudenberg2016whither}, 2) the model of GT should be efficient with limited complexity  \cite{bacci2015understanding}, and 3) the prediction capability of GT could be improved using learning theory \cite{fudenberg2016whither}

\section{Conclusion}
\label{sec_conclusion}

This paper provides a summary of recent developments of GT applied in VNs, discussing the applications of GT in the existing VNs and the future vehicular communications integrated with 5G mobile network technologies. After briefly introducing the concepts of relevant terms of GT, the current game theoretic-approaches used to solve the challenges in VNs are comprehensively surveyed from the perspectives of QoS (e.g., routing selection, power control, and spectrum resources allocation) and security (e.g., intrusion detection, privacy-preserving, and trust management). Furthermore, we survey the works of GT employed in next-generation VNs integrating with emerging 5G technologies such as edge computing, heterogeneous networks, UAV-assisted communications, and SDN technologies. Finally, the remaining challenges and the corresponding future researches applying GT in VNs are discussed.

 \section*{Acknowledgment}
	This work was supported by the National Nature Science Foundation [61373123, 61572229, U1564211, and 6171101066]; Jilin Provincial Science and Technology Development Foundation [20170204074GX, 20180201068GX] and Jilin Provincial International Cooperation Foundation [20180414015GH].
	
	\bibliographystyle{IEEEtran}
	\bibliography{references.bib}
	
\end{document}